# Market Crowd's Trading Behaviors, Agreement Prices, and the Implications of Trading Volume

Previous Title Was

Market Crowd's Trading Conditioning, Agreement Price, and Volume Implications

Leilei Shi*[1], Bing Han[2], Yingzi Zhu[3], Liyan Han[4], Yiwen Wang[5], and Yan Piao[1]

[1] Department of Modern Physics, University of Science and Technology of China (USTC), China
[2] The Rotman School of Management, University of Toronto, Canada
[3] Schools of Economics and Management, Tsinghua University, China
[4] Schools of Economics and Management, Beihang University, China
[5] Huashang Fund Management Co., Ltd., China

This Draft is on December 30, 2013
The First Draft is on January 31, 2012

(Comments welcome)

*Corresponding author works in a prestigious securities company in China. His e-mail address: Shileilei8@aliyun.com or leilei.shi@hotmail.com

The authors thank valuable discussions with Huaiyu Wang, Martin Schaden, Binghong Wang, Hongkui Yu, Yannick Malevergne, Stephen Figlewski, Pengjie Gao, Lei Lu, Yonggan Zhao, Ding Chen, Chengling Gou, Wei Xiong, Jie Hu, Lifang Gu, Yonghong An, Mingshan Zhou, Yu-En Lin, Domenico Tarzia, and participants at 2011 Annual Meeting of the Midwest Finance Association (USA), 2013 China Meeting of the Econometric Society, the 6[th] (2013) China Finance Review International Conference, STATPHYS25 (25[th] IUPAP International Conference on Statistical Physics), the 10[th] (2013) Chinese Finance Annual Meeting, and the World Finance & Banking Symposium (2013) in Beijing etc. Shi especially appreciates an interview to discuss the paper by Andy Webb from *Automated Trader Magazine*. Of course, we are responsible for all remaining errors and omissions.

# Market Crowd's Trading Behaviors, Agreement Prices, and the Implications of Trading Volume

(December 30, 2013)

## Abstract

It has been long that literature in financial academics focuses mainly on price and return but much less on trading volume. In the past twenty years, it has already linked both price and trading volume to economic fundamentals, and explored the behavioral implications of trading volume such as investor's attitude toward risks, overconfidence, disagreement, and attention etc. However, what is surprising is how little we really know about trading volume. Here we show that trading volume probability represents the frequency of market crowd's trading action in terms of behavior analysis, and test two adaptive hypotheses relevant to the volume uncertainty associated with price in China stock market. The empirical work reveals that market crowd trade a stock in efficient adaptation except for simple heuristics, gradually tend to achieve agreement on an outcome or an asset price widely on a trading day, and generate such a stationary equilibrium price very often in interaction and competition among themselves no matter whether it is highly overestimated or underestimated. This suggests that asset prices include not only a fundamental value but also private information, speculative, sentiment, attention, gamble, and entertainment values etc. Moreover, market crowd adapt to gain and loss by trading volume increase or decrease significantly in interaction with environment in any two consecutive trading days. Our results demonstrate how interaction between information and news, the trading action, and return outcomes in the three-term feedback loop produces excessive trading volume which includes various internal and external causes. Finally, we reconcile market dynamics and crowd's trading behaviors in a unified framework by Shi's price-volume differential equation in stock market where, we assume, investors derive a liquidity utility expressed in terms of trading wealth which is equal to the sum of a probability weighting utility and a reversal utility in reference to an outcome.

**Key words:** trading volume, crowd's behaviors, adaptive hypothesis, asset price(s), differential equation, behavior analysis, prospect theory, liquidity utility

JEL Classifications: G12, G02, D83

# 市场群体的交易行为、认同价格以及交易量的内涵

(2013 年 12 月 30 日)

## 摘要
**(Abstract in Chinese)**

长期以来，金融学术领域里的文献只注重价格和收益率，却较少研究交易量。在最近的二十年里，金融学术文献已经开始研究价格和交易量两者与经济基本量之间的相互关系，并且探讨交易量的行为内涵，例如投资者对风险的态度、过度自信、不同观点以及关注程度等等。然而，我们还是对交易量的认识知之其少。本文根据行为分析，用交易量概率来表示市场群体的交易频率，并且通过我国股市来实证检验涉及交易量与价格之间不确定关系的两种适应性假说。实证结果表明：市场群体在每日交易的时间窗口内除了采用简单的经验法则之外，同时还采用有效的适应性方式来从事股票交易，并且逐步倾向于形成一个结果和认同的资产价格；无论该资产价格是否明显地被高估或低估，市场群体在相互作用和竞争的过程中往往能够形成这样一个稳态的均衡价格。这表明了资产价格不仅包含了基本价值同时还包含了非公开信息、投机、情绪、关注、赌博和娱乐等价值。此外，在任意两个连续交易日之间，市场群体在与市场环境的相互作用过程中，通过交易量的增加或减少来有效地适应盈亏。我们的研究结果说明了在由信息、交易与收益结果三项构成的反馈环中，它们之间的相互作用是如何导致了过度交易的，这其中包含了导致过度交易的各种内外因素。最后，我们假设股票市场中的投资者是通过交易财富来产生流动性效用，它等于概率加权效用与相对于结果为参照系的反转效用之和，从而推导出 Shi 氏价-量微分方程，将市场动力学行为与群体交易行为协调在一个统一的框架体系。

**关键词**：交易量，群体行为，适应性假说，资产定价，微分方程，行为分析，前景理论，流动性效用
JEL 分类：G12，G02，D83

## 1. INTRODUCTION

It has been long that literature in finance focuses mainly on price and return behavior but much less even completely ignoring on trading volume. There is no trading volume in neoclassical finance models, for examples, CAPM (Sharpe, 1964), price Brownian motion (Samuelson, 1965), option pricing model (Black and Scholes, 1973), efficient market hypothesis—EMH (Fama, 1970), and arbitrage pricing theory (Ross, 1976) etc. However, trading volume is very high on the world's stock market, contrary to the prediction of rational models of investment in a neoclassical paradigm (Barberis and Thaler, 2003). What is surprising is how little we really know about trading volume (Lee and Swaminathan, 2000).

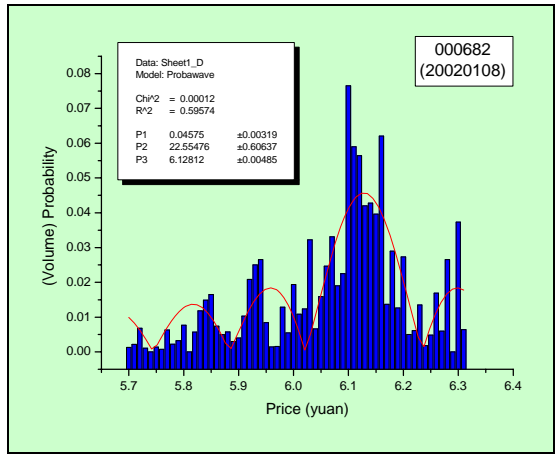

Figure 1: A stock price and volume historical data on a trading day and its test results

Trading volume distributes over a price range in a limited number of patterns on a trading day in stock market (Shi, 2006). So, it is reasonable that the market behaviors could be described in price and volume coordinates (For an illustration of this, see Figure 1). It might be governed by the price-volume probability differential equation that is derived from a liquidity utility hypothesis[1] in an economy where traders abide by a trading rule—"price first and time first", i.e., an actual trading price path in a variety of options is selected by the least trading price or cost variation (Shi, 2006). Shi (2006) empirically tests trading volume distribution over a price range on a trading day by two sets of explicit eigenfunctions—analytical and closed form solutions from the price-volume probability wave equation (see a test example in Figure 1). It is a new method in mathematical finance. However, the implications in the differential equation have not yet been understood for economists and finance people, for examples, trading volume probability, stationary equilibrium[2], liquidity utility hypothesis, trading conditioning (Shi et al., 2010), and the least trading price variation to generate optimal choice and preference structure in interactive and competitive trading action etc. We explain them and thus form the subject of the present study.

---
[1] It is called as a transaction energy hypothesis before (Shi, 2006).
[2] See its definition in Section 3.2 in this paper.

Market crowd are all traders who interact among themselves with a variety of heterogeneous beliefs, preferences, and biases. Based on information and news, they have expectation on return and trade accordingly in stock market. Some are rational and trade in terms of stock fundamental value (Samuelson, 1937; Von Neumann and Morgenstern, 1944), others are irrational and act in gamble and entertainment (Hirshleifer, Subrahmanyam, and Titman, 2006; Kumar, 2009; Dorn and Sengmueller, 2009; Hoffmann and Shefrin, 2012), and the others between the two groups are boundedly rational and behave adaptive, goal-oriented, and intended rationality (Simon, 1956; Conlisk, 1996; Jones, 1999; Kahneman, 2003; Rachlin, 2003; Gabaix et al., 2006). It is stock market, a consequence of trading action in a feedback loop between market crowd and market environment in stock market.

In the past 20 years, academics have already linked both price and volume to economic fundamentals (He and Wang, 1995; Wang, 2002) and paid increasingly attention to the information contained in trading volume. In neoclassical framework, Lo and Wang (2006) derive an intertemporal asset pricing model of multiple assets in the spirit of ICAPM (Merton, 1973), in which investor's attitude toward risks can be extracted from trading volume.

In newly emerging behavioral finance, scholars associate trading volume with the degree of heterogeneity in investors' beliefs, preferences, and biases. Lee and Swaminathan (2000) show that past trading volume provides an important link between "momentum" and "value" strategies. It helps to reconcile intermediate horizon "underreaction" and long horizon "overreaction" effects.

Moreover, Odean (1998b) puts forward the hypothesis that overconfidence produces excessive trading in stock market. Odean (1999) explains why those actively trade in financial markets to be more overconfident than general population by three reasons: selection bias, survivorship bias, and unrealistic belief, and tests overconfident trading hypothesis by investigating whether the trading profits of discount brokerage customers are sufficient to cover their trading costs. Barber and Odean (2000) show that active trading results in poor performance and is hazardous to wealth. Barber et al. (2009) figure out that individual investor trading behavior results in systematic and economically large losses, about an annual performance penalty of 3.8 percentage points in Taiwan stock market. Such irrational behavior can be explained by overconfidence. Grinblatt and Keloharju (2009) document that both overconfident investors and sensation-seeking investors trade more frequently. Tested the trading volume predictions of formal overconfidence models, Statman, Thorley, and Vorkink (2006) find that share turnover is positively related to lagged returns in both market-wide and individual security for many months. It is interpreted as the evidence of investor overconfidence and disposition effect. Ben-David and Doukas (2006) present evidence about the mechanism that links investor overconfidence to disposition effect. They find that overconfidence heightens the trading frequency of institutional investors while impacting asymmetrically on their trading between past winners and losers. In addition, Graham, Harvey, and Huang (2009) find that investors who feel competent trade more, and thus overconfidence leads to higher trading frequency.

Glaser and Weber (2007) consider three forms of overconfidence in psychology: miscalibration, volatility estimates, and the better than average effect. Miscalibration is the overestimating the precision of information about the value of a financial security (Graham, Harvey, and Huang, 2009). The degree of overconfidence in judgment in the form of miscalibration is the tendency to overestimate the precision of one's information (Biais et al., 2005). By testing correlations between the degree of overconfidence and trading volume, however, they find that overconfidence as measured by calibration questions or miscalibration scores is not significantly related to trading volume, although investors who think that they are above average trade more. It is consistent with the similar findings by Biais et al. (2005) and Dorn and Huberman (2005). Thus, they conclude that the overconfident hypothesis is too broad, we should specify what type of overconfidence may be influencing trading behavior and treat it with caution, given that the "differences of opinion" literature better explains high levels of trading volume when compared to the "overconfidence" literature. Hong and Stein (2007) go further: trading volume appears to be an indicator of sentiment. In other words, when prices look to be high relative to fundamental value, disagreement on price is strong, and volume is abnormally high. Thus, they propose a disagreement model which has direct implications for the joint behavior of stock price and trading volume.

Barber and Odean (2008) also have an alternative behavioral explanation about what drives trading volume: volume is more directly related to actual attention and capture the change in investor attention to a stock by its abnormal daily trading volume. Hereafter, Huo, Peng, and Xiong (2009) use trading volume as a proxy of investor attention to study overreaction and underreaction on the price and earnings momentum profit. Yu (2012) finds that high attention causes individual investors to reduce their stock holdings dramatically or trade more frequently when the market level is high, and to reduce their stock holdings modestly or trade less frequently otherwise. Da, Engelberg, and Gao (2011) propose a direct measure of investor attention using the frequency of crowd's search in Google and then revisit the relation between investor attention and asset prices via the frequency.

Besides, Barber, Odean, and Zhu (2009) explain that psychological biases and decision makings by heuristics likely lead to the correlated trading of individuals with unusually high trading volume.

Therefore, there is a long list of personal characteristics, psychological biases, and physiological causes that are attributed to excessive trading volume, e.g. attitude to risks, overreaction, overconfidence, sensation-seeking, disagreement, attention, sentiment (Han, 2007), gender (Barber and Odean, 2001), and intuition (Hoffmann and Shefrin, 2012) etc. These studies have yielded some success. However, there are seemly competing behavioral explanations on trading volume (Barberis and Thaler, 2003)[3]. It inspires us to explore a unified explanation.

In behavior analysis (Coon and Mitterer, 2007; Pierce and Cheney, 2004), we assume that all behavior is the product of two kinds of variables: biological (internal) and environmental (external)[4]. Except for the internal, covert, and biological elements and factors of excessive trading volume [5], a long list of external, overt, and environmental elements and factors, from which market crowd infer and predict the probability of gain and loss, elicit market crowd's response and cause them to trade in risky choice, for examples, dividend (Shiller, 1981), size and book-to-market ratio (Fama and French, 1995), economic fundamentals (He and Wang, 1995), private information (Amihud, Mendelson, and Pedersen, 2005; Chordia, Roll, and Subrahmanyam, 2008), news announcement (Jia, Wang, and Xiong, 2013), a technical analysis cue (Lo, Mamaysky, and Wang, 2000; Zhu and Zhou, 2009), entertainment (Dorn and Sengmueller, 2009; Hoffmann and Shefrin, 2012), social interaction (Shiller, 1984 and 1995; Han and Yang, 2013; Han and Hirshlerfer, 2013), linguistic diversity (Chang, Hong, Tiedens, and Zhao, 2013), and so on. Past return, reference price effect, tax-loss selling, and the size of the holding period capital gain or loss etc affect trading frequency (Grinblatt and Keloharju, 2001; Griffin, Nardari, and Stulz, 2006; Shi et al., 2010). According to prospect theory (Kahneman and Tversky, 1979), investors tend to frame investment choices in terms of gain and loss, the outcome of prior trading action. Disposition sellers are risk-averse over gains and risk-seeking over losses (Barberis and Xiong, 2009). They adapt to sell winners too early and hold losers too long in stock market (Shefrin and Statman, 1985; Grinblatt and Han, 2005; Dhar and Zhu, 2006).

---

[3] Barberis and Thaler (2003) assert that there are obviously competing behavioral explanations for some of the same empirical facts in behavioral financial studies. We are still much closer to the beginning of the research agenda than we are to the end, although there is a lot of accomplishment in a short period of time.

Adaptive learning is seen everywhere in economics and finance, for examples, adaptive to a stock fundamental value in neoclassical paradigm (Fama, 1970; Ross, 1976) or adaptive to an outcome in behavioral framework (Simon, 1956; Kahneman and Tversky, 1979; Lux 1995; Jones, 1999). Lo (2004, 2005, and 2011a) attempt to reconcile rational hypothesis in neoclassical finance and boundedly rational hypothesis in behavioral finance using a descriptive and qualitative Adaptive Markets Hypothesis (AMH). Neely, Weller, and Ulrich (2009) find that foreign exchange market supports AMH (Lo, 2004), but not EMH (Fama, 1970). Based on prospect theory (Kahneman and Tversky, 1979), Arkes et al (2008) review several candidates suggested for a reference point and use two approaches to test adaptive behavior in terms of a shift of reference point in the direction of a realized outcome: questionnaire experiments and a stock trading game in laboratory. Barberis and Xiong (2012) presents a model to explain the greater turnover in rising markets and the heavy trading of highly valued assets in crowd's adaptation in which investors may derive utility from realizing gains and losses. Based on how prior outcomes affect risky choice (Kahneman and Tversky, 1979), moreover, Barberis, Huang, and Santos (2001) study asset prices in the assumption that investors derive a direct utility not only from consumption but also from change in the value of their financial wealth. These empirical evidences and non-expected utility models have made substantial progress.

---

[4] Behavior analysis is one of major branches in psychology. Its major organization is Association for Behavior Analysis International ( http://www.abainternational.org/ ). Behavior analysis and cognitive psychology are intertwined and reconcilable. The findings and even the theories by cognitive psychology may be interpreted in terms of reinforcement and punishment acting on observable behavioral patterns (Rachlin, 2003) because of interaction between organism and environment.

[5] Internal, covert, and biological variables can be viewed as part of an organism's environment that is rooted in biology, which produce comfort, end discomfort, or fill an immediate physical need for behavior selection etc.

However, it is still not understood how efficiently market crowd choose a reference point and achieve agreement on an outcome or an asset price in trading action, what observable behavior they take for outcome adaptation in stock market, and why trading volume or frequency can represent subjective behaviors such as attitude toward risks (Lo and Wang, 2006), overconfidence (Odean, 1998b), disagreement (Hong and Stein, 2007), and attention (Barber and Odean, 2008; Huo, Peng, and Xiong, 2009; Da, Engelberg, and Gao, 2011; Yuan, 2012) etc.

We extend Shi's price-volume differential equation (Shi, 2006), study the volume uncertainty associated with price, and attempt to solve these questions. We incorporate subjective thinking into the equation, measure the frequency of market crowd's trading action by trading volume probability in terms of principles in behavior analysis (Shi et al., 2010), and test two kinds of adaptive trading hypotheses relevant to the volume uncertainty associated with price. They are: 1) market crowd's adaptive behavior in reference to an outcome, an agreement price, or a stationary equilibrium price on a trading day[6]; and 2) market crowd's adaptive behavior to gain and loss by trading volume increase or decrease in any two consecutive trading days.

Like the non-expected utility models by Barberis, Huang, and Santos (2001) and Barberis and Xiong (2012), we also consider trading outcome and change in financial wealth in our model. What difference is that we do not modify investor's expected utility in a traditional framework. We model the market populated by momentum traders, reversal traders, and the interactive traders who are influenced by momentum and reversal traders on a trading day[7]. They trade a stock in adaptive to an outcome on a daily basis. We reconcile market dynamics and crowd's trading behaviors in a unified framework by Shi's price-volume differential equation in stock market where investors derive a liquidity utility expressed in terms of trading wealth, which is equal to the sum of a volume probability weighting utility and a reversal utility in reference to a stationary equilibrium price—the outcome of crowd's trading action (Shi, 2006)[8]. It is a new method in mathematical finance, from which we obtain explicit models and the extremal functions that describe market crowd's optimal choice over a price range in interactive and competitive trading action on a trading day.

The remainder of the paper is organized as follows: section 2 introduces measurement in behavior analysis, proposes a notion of the trading action in stock market according to behavior analysis, and measures the frequency and preference of market crowd's trading action; section 3 explains two trading behavioral hypotheses relevant to the volume uncertainty associated with price; section 4 is empirical test. We establish a liquidity utility hypothesis and derive a price-volume probability differential equation in section 5. Moreover, we have discussions in section 6. Final are conclusions, contributions, and extensions.

---

[6] A stationary equilibrium price is the price on a trading day at which the corresponding volume is the maximal (Shi, 2006).
[7] Financial academics has repeatedly documented price momentum in a short term but reversal in a long term (Shiller, 1981; Jegadeesh and Titman, 1993; Hong and Stein, 1999; Lee and Swaminathan, 2000; Grinblatt and Han, 2005).
[8] Intuitive predictions by a judgmental heuristic—representativeness, which lead cognitive biases from time to time, are insensitive to the prior probability of the outcome (Kahneman and Tversky, 1973; Tversky and Kahneman, 1974). Thus, price fluctuation in relative to outcome is the other important aspect we should study in decision making. It could be explained by reinforcement and punishment in behavior analysis.

## 2. MARKET CROWD'S TRADING FREQUENCY AND ITS MEASUREMENT

As we study the relationship between liquidity increment or injection, any a stationary equilibrium price jump, and change in trading volume by a price-volume probability wave equation (Shi, 2006), we have to explore the volume uncertainty associated with the jump, for example, there are both positive and negative correlations between change in volume and jump in a stationary equilibrium price or mean return. We therefore attempt to incorporate subjective thinking into the equation and apply principles in behavior analysis and prospect theory into our study.

### 2.1 Measurement in Behavior Analysis and a Notion of the Trading Action

Behavior analysis is a comprehensive approach to the study of organism's behavior by observable variables which have been emphasized since Watson (1913). It has a strong focus on environment-behavior relationships, that is, how organisms adapt their behavior to meet the ever-changing demands of the environment (Pierce and Cheney, 2004; Staddon, 2010). Its theory could provide a scientific account of the learning and behavior of traders in terms of outcomes in stock market.

Pavlov (1904) discovers conditioned reflex or classical conditioning. It is a physiological response to expect that a physiological unconditioned stimulus will follow whenever a conditioned stimulus is present. Thorndike (1913), a pioneer in operant conditioning, studies the trial-and-error learning that is based on unobservable states of mind. Skinner (1935) makes the distinction between two types of conditioned reflex, the difference between operant and respondent behavior, based the operant conditioning on Darwin's principle of selection and evolution. Whereas respondent conditioning directs attention to environmental events and behavioral responses, his operant conditioning involves adaptive behavior by its consequences (Skinner, 1938). The outcomes that follow behavior determine the increase or decrease probability of current response and action (Coon and Mitterer, 2007).

Like classical conditioning, operant learning is also based on information and expectancy. In the presence of a discriminative stimulus[9], reinforcement will occur if and only if the operant response occurs (Dragoi, 1997). Skinner calls this relationship a three-term contingency (Pierce and Cheney, 2004): in a given situation or setting (discriminative stimulus), an organism takes action (operant class) to produce an effect or consequence (reinforcement). The current frequency of operant is explained by relationship between operant and reinforcement in the past. Today, operant rate or frequency is considered a basic, convenient, and better measure for subjective thinking and preference than any other observable behaviors in learning. For example, saliva volume is hardly measured in experiment.

We apply principles in behavior analysis to real-world in financial market and propose a notion of the trading action[10].

---

[9] A discriminative stimulus is an event that precedes an operant and sets the occasion for behavior (Pierce and Cheney, 2004).

[10] The trading action is called as trading conditioning before (Shi et al, 2010) which is borrowed from operant conditioning in behavior analysis. Trading conditioning is foreign to economists and finance people. So, we use the

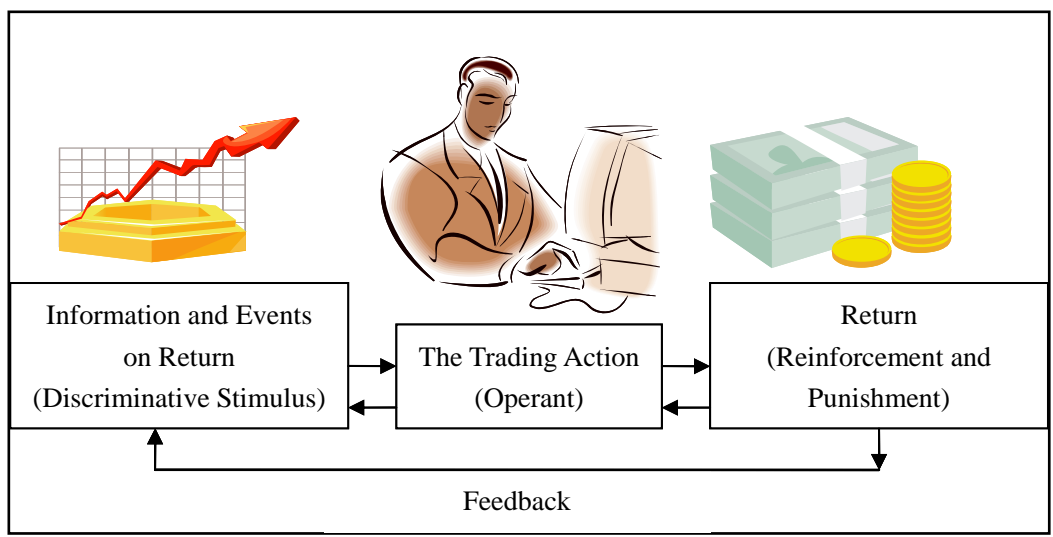

Figure 2: Three-term relationship in a feedback loop in the trading action

The trading action is a consequence of both classical and operant conditioning. One reason for the interrelationship of classical conditioning and operant conditioning is whatever a participant trades is applied to the whole neurophysiologic system that is usually showed by emotion, sentiment, and other physiological and psychological behaviors etc., not just trading alone (Pierce and Cheney, 2004; Lo, 2011b)[11]. It is intentional, adaptive, and goal-oriented behavior in learning, not simply and passively respondent one. Its discriminative stimulus classes are information and news on return, its operant in response to them with expectation on return is the trading action that pairs shares, and its reinforcement and punishment are return, in which positive return or gain is reinforcement for stock holders while negative return or loss is punishment. Obviously, the return feedback is a part of information and news, but there is no price volatility or return at all if there is no trading action. Information and news cannot result in any price volatility or market return if there is neither trading nor dividend. Moreover, people are the major source of information and news, and past return—the outcome of prior trading—regulates the frequency of current trading action. Therefore, there are reciprocal causations in the three-term relationship, according to social cognitive and learning theory (Bandura, 1986 and 2001). See Figure 2 and read Section 2.2.4.

## 2.2 Decision Making (Choice) by Trading Action

In this subsection, we first give underlying hypotheses relating to traders in stock market, clarify the causality between the trading action (trading liquidity) and the joint behavior of price and volume, and measure subjective thinking by trading volume probability. We define the trading action in terms of behavior analysis.

---

trading action instead of trading conditioning in this paper.
[11] Lo (2011) explains from neurosciences that our most fundamental reactions to monetary gain and loss such as fear and greed are hardwired into human physiology.

2.2.1 Underlying hypotheses in the trading action

There are several underlying hypotheses relating to traders in stock market when we study trading behaviors using the principles in behavior analysis. First, we assume that intangible and subjective behaviors such as attitude to risks (Lo and Wang, 2006), overconfidence (Odean, 1998b), disagreement (Hong and Stein, 2007), attention (Barber and Odean, 2008; Huo, Peng, and Xiong, 2009; Da, Engelberg, and Gao, 2011), and entertainment (Dorn and Sengmueller, 2009) can be represented to a large extent by external and intrinsically observable behavioral patterns such as trading volume and search frequency[12]. Second, there is interaction between traders and market environment in a feedback loop. It is best understood in terms of an interaction over time between the frequency of the trading action and its prior consequences—market return (Rachlin, 1999). Third, we assume that a simple behavior in learning, for examples, the behaviors of plants, single-celled animals, and rats, is one of elements of a complex one such as trading behavior in stock market. It allows us to study trading behaviors using principles in behavior analysis. Last, investors interact and compete among themselves rather than being independent in stock market.

2.2.2 Two kinds of explanation

In science, we talk about two kinds of causation: immediate and remote (Pierce and Cheney, 2004; Staddon, 2010). Immediate causation is the kind of mechanism studied in physics and chemistry, for examples, Newton's laws, Maxwell's equations, and Schrŏdinger's equation etc. It is a direct cause-and-effect explanation. In the study of the trading action, an immediate explanation might refer to the physiology and psychology of a trader, for examples, gender (Barber and Odean, 2001), overconfidence (Odean, 1998b), and attention (Barber and Odean, 2008). In contrast, remote causation is the kind of sciences studied in evolutionary biology, geology, astronomy, and artificial intelligence etc, for example, the principle of selection in terms of outcomes. It is an indirect cause-and-effect or functional explanation. In the study of the trading action, a functional explanation might refer to not only the information and news—heuristics and representativeness, but also the outcomes of prior trading—adaptation by return reinforcement or punishment (Tversky and Kahneman, 1974; Staddon, 2010; Shi et al., 2010). The former functional explanation is classified into template learning, and the latter is called as reinforced learning (Staddon, 2010). Both immediate and remote causal explanations are acceptable in science. Ultimately, both types of explanation will provide a more complete account of trading behaviors in learning (reference to Figure 2).

2.2.3 Liquidity, price, and volume

---

[12] It is similar to invisible gravity that can be measured by the product of a body's mass (observable) and acceleration (measurable) in physics, a consequence of interaction between the body and the Earth (environment).

In order to understand trading behaviors, let us establish a direct cause-and-effect relationship between the trading action that pairs shares, price change, and volume in stock market. Trading action produces liquidity. The liquidity is the amount traded in a time interval.

There is no price change at all if there is no trading action that pairs shares in stock market, i.e., only the trading action results in price change, the consequence of prior trading action. It produces a positive return if demand volume is larger than supply volume, a negative return if demand volume is less than supply volume, and no return if demand volume is equal to supply volume in pairings at a price in a time interval, respectively. In addition, if there is trading action, then, there is change in volume in a time interval, and vice versa. Therefore, although there is a long list of causes that could affect price and volume, but the final cause is one—only the trading action or trading liquidity results in change in price and volume in stock market. From the immediate causation between the trading action (trading liquidity) and the change in price and volume, Shi (2006) derives the price-volume probability differential equation that governs the volume distribution over a price range with liquidity constraints by two sets of eigenfunctions (see Figure 1). That is, there is mechanistic relation between the trading action (trading liquidity) and the change in price and volume, like an equation or law in physics. We will derive Shi's equation in terms of supply-demand law, behavior analysis, prospect theory in section 5.

In addition, it is remote causation in a feedback loop that market gain and loss, the outcome of prior trading action, increases or decreases the frequency of the trading action. We can most simply describe the frequency in response to price change or market return by trading volume probability in the price-volume differential equation. It is a functional and descriptive explanation. The feedback provides the conceptual link between explanations in terms of physiological and psychological causes and explanations in terms of goals and final outcomes (see Figure 2).

Now, let us consider a rigid trading system, in which we assume that there is no subjective thinking on price change. In another word, gain and loss has no reinforcement and punishment value for market crowd at all. Total trading volume in a time interval is the same at any same time interval when price changes. If large supply-demand imbalance breaks stationary equilibrium and results in a stationary equilibrium price jump, then there is one-to-one correspondence between liquidity increment and jump because total volume is the same in two time intervals right before and just after jump.

However, this assumption is not true in real world. Market crowd do respond to gain and loss in stock market (Kahneman and Tversky, 1979; Shefrin and Statman, 1985; Barberis and Xiong, 2009). They are sensitive to and adapt to gain and loss, a consequence of prior trading action, by trading volume increase or decrease significantly in market evolution (Shi et al, 2010).

2.2.4 Measurement on belief, the trading action, and preference

A trader buys or sells a stock because he accepts its price and believes that it is worth expecting reward or avoiding punishment in the face of certain risky choice after his judgment on its value. To meet a trader's own satisfying feelings (Simon, 1956), a stock trading value includes not only fundamental value but also private information value (Kyle, 1985), speculative value (Hirshleifer, Subrahmanyam, and Titman, 2006; Han and Kumar, 2013), sentiment value (Barberis, Shleifer, and Vishny, 1998; Baker and Wurgler, 2006; Han, 2007), attention value (Da, Engelberg, and Gao, 2011), gamble value (Kumar, 2009), and entertainment value (Dorn and Sengmueller, 2009; Hoffmann and Shefrin, 2012) etc. We could measure one's belief on a stock value by his trading price. In addition, investor's trading creates liquidity, and liquidity facilitates trading. We could measure the trading action by trading liquidity.

It is a fact that people have physiological demand for clothes, foods, and services in life. The physiological response to those is a kind of unconditioned response. Printed money obviously has little or no value of its own at all, neither eaten nor drunk. However, when we associate money, asset, and return with the necessities of life and services tightly through exchange in commodity exchange economy, we do produce the same physiological response to them. Money, asset, and return are the most important source of economic reinforcement (Pierce and Cheney, 2004).

Stock holders experience increase or decrease in wealth if price rises or drops in stock market. It is a kind of conditional response if they have the same physiological response, for example, fear and greed, as they do for the necessities of life and services. It is a kind of conditional action if they trade to approach gain and avoid loss after they analyze, infer, and make a decision in the process of information and news[13]. The trading action is the inter-relative behavior of both conditional response and conditional action (see Figure 2).

There are several characteristics in the trading action. First, return does not lose reinforcement and punishment value because money is not too much to lose its reinforcement value in commodity exchange economy. Second, return has both good and bad quality in stock market. It has reinforcement and punishment value simultaneously. Positive return is reinforcement for stock holders, but punishment for cash or short holders. Negative return is punishment for stock holders, but reinforcement for cash or short holders. One buys if he expects stock price rise in the future. Otherwise, one sells or shorts if he believes its drop. Third, gain or loss occurs in an uncertain time after one buys in stock market. It makes tremendous resistance to stop trading, similar to partial reinforcement in behavior analysis. Forth, when one must cope with return uncertainty in stock market, he infers the probability of gain and loss from the most likely cause(s) of return, takes the trading action, and adapts himself to gain and loss with a higher or lower trading probability. The return feedback could influence his emotion and sentiment by surprise and the violation of expectation, affect his judgment, and change his expectation on return[14]. For example, a stop loss trader who sells a loser just before is likely to adapt and buy it again at a higher price at the moment when he is lured by price bounce. The prior punishment does not reduce the frequency of the trading action. This effect is a special case of the innate tendency to seek gain and avoid loss in response to certain information and news without being reinforced by market reward and punishment[15]. Therefore, one could adapt to trade more frequently, based on both antecedent causes and return outcomes.

---

[13] Soros (1987) attempts to explain market crowd's trading behaviors by a reflexivity theory.
[14] A generalized reinforcement could be praise, gamble, or entertainment etc.

We define the trading action as: in the presence of information and news, one responds to them and trades with expectation on future return after he analyzes a stock value, infers the probability of gain and loss, and makes a decision; one learns from gain or loss by which reinforcement or punishment feedback could influence his emotion, affect his judgment, and alter his expectation on return, adapts himself in evolution, and increases or decreases his trading frequency (see Figure 2).

Staddon (2010) classifies learning into two main types—template learning and reinforced learning. So, the trading action could be distinguished in two parts: the trading action in terms of information and news—heuristics and representativeness (template learning), and the trading action in terms of outcome—adaptation to an outcome, an agreement price, or a stationary equilibrium price (reinforced learning). Both heuristics and adaptation are insensitive each other (Kahneman and Tversky, 1973; Tversky and Kahneman, 1974).

The more frequently the investor trades, the stronger expectation on return he has in response to relevant information and news, and vice versa. We can measure the intensity of one's response or trading action by his trading frequency or volume probability. Moreover, an investor prefers to trade more at one price than another over an acceptable price range in a given time interval. The higher frequency or volume he trades at a price over a price range, the stronger preference he shows in his trading action. Thus, trading frequency or volume represents his trading preference strength.

## 2.3 Measurement on Market Crowd's Trading Frequency and Preference

We measure market crowd's trading action by liquidity and their trading frequency by trading volume probability, respectively[16]. The frequency represents the intensity of both their trading action and their trading preference at a price over a price range in a time interval.

Market crowd are a group of rational, boundedly rational, and irrational traders. We measure them by total trading volume rather than the number of traders. A unit of market crowd is equivalent to a unit of trading volume, a share. For example, an individual investor who buys 1,000 shares is a representative of 1,000 units of market crowd. Market crowd made up of a few institutional investors is greater than those consisting of a larger number of individual investors, if the former trade more volume. A few institutional investors may have much stronger impact on price than a larger number of individual investors (Nofsinger and Sias, 1999).

Based on information and news, market crowd respond to them and trade with expectation on return. The larger the trading volume is at a price over a trading price range in a time interval, the larger volume probability the market crowd trade at the price. Thus, the trading frequency is higher at the price. It is stronger that the intensity of market crowd's trading preference at the price, and vice versa. Therefore, we can measure both the frequency of market crowd's trading action and the intensity of their trading preference at a price by trading volume probability in Shi's price-volume differential equation (Shi, 2006).

---

[15] It is called autoshaping in behavior analysis (Brown and Jenkins, 1968). In autoshaping, stimuli that predict something of value gain control over behavior, i.e. certain kinds of situation automatically entail certain kinds of action (Staddon, 2010).

[16] In our study, total daily trading volume is about 360,000,000 shares in average which is much greater than any a tick by tick trading volume.. Therefore, we can use accumulative trading volume probability to represent the frequency of market crowd trading action, according to probability and statistics.

Here, crowd's preference refers to trade a stock differently over an acceptable price range in terms of not only information and news but also outcomes, a process which could generate a crowd's optimal choice in competing for limited resources as a whole to meet the degree of their heterogeneous satisfaction.

## 3. THE VOLUME UNCERTAINTY ASSOCIATED WITH PRICE

There are two kinds of volume uncertainty with price. One is that trading volume distribution over a price range on a trading day (Shi, 2006), and the other is that current total trading volume with a past mean return or jump in a stationary equilibrium price in any two consecutive trading days[17]. They are explained by two kinds of adaptive trading behaviors in behavioral framework, respectively: adaptation to an outcome, a stationary equilibrium price, or an asset price (with a mean return of zero or very close to zero), and adaptation to gain and loss, the outcome of prior trading action (with a significant past mean return).

### 3.1 Two Trading Behavioral Hypotheses

We put forward two trading behavioral hypotheses relevant to the volume uncertainty associated with price and explain them in terms of a supply and demand law, behavior analysis (Pierce and Cheney, 2004; Staddon, 2010), and prospect theory (Kahneman and Tversky, 1979).

**Hypothesis One (An Estimated Stock Value Hypothesis):** Market crowd trade a stock in simple heuristics[18] and efficient adaptation[19], gradually tend to generate an outcome or a stationary equilibrium price in interaction and competition among themselves, and form a common belief (agreement) on a stock value (asset price) on a trading day.

**Hypothesis Two (An Adaptive Trading Hypothesis):** Market crowd adapt to gain and loss by trading volume increase or decrease efficiently in interaction with environment in market evolution in any two consecutive trading days.

### 3.2 Trading Volume Distribution and Stationary Equilibrium

---

[17] Extensive papers have showed that there is an uncertain relation between trading volume and stock return, though there is less agreement on its explanation (Wang, 1994; Lee and Swaminathan, 2000; Llorente et al., 2002).
[18] Judgment in heuristics which leads cognitive biases has been explained by Tversky and Kahneman (1974).
[19] It specifies the adaptation that market crowd trade a stock in reference to an outcome or a stationary equilibrium price rather than its fundamental value that is hardly calculated. We will test it in this paper.

As we see in Figure 1, trading volume distributes over a trading price range in a limited number of patterns on a trading day (Shi, 2006). We are going to explain it.

There are three typical reversal, restoring, and negative feedback forces in equilibrium. In the first case is gravity; in the second is charge or spring; and the third, competition among individuals for limited resources (Staddon, 2010). It belongs to the third category about stationary equilibrium in stock market.

Market crowd are all traders with a variety of heterogeneous beliefs, preferences, and biases in stock market where the vast numbers of separate and seemingly independent individuals buy and sell a stock. They trade a stock because they accept its price and believe that it is worth expecting gain (reward) or avoiding loss (punishment) in the face of certain risky choice after his judgment on its value. In addition, investors usually trade in heuristics (Kahneman and Tversky, 1973; Tversky and Kahneman 1974) and adaptation (Shi et al., 2010) to reduce cost and speed up the process of finding its value. They trade a stock not at an exclusive price but over a price range because of free competition for limited resources among themselves. Total trading wealth distributes over a price range.

If individuals make judgment independently, they often violate consistency and coherence because of decision frame (Tversky and Kahneman, 1981). Everyone prefers to buy a stock at a price as low as possible or to sell it as high as possible. However, whereas one waits for a while to buy (sell) a stock at a lower (higher) price, he also worries that he has to buy (sell) it at a higher (lower) price later because of free competition from others. When people trade a stock on a trading day, they interact and compete among themselves, and achieve agreement on an outcome or a stationary equilibrium price widely (Shi, 2006). There are a limited number of trading volume distribution patterns over a price range on a trading day (Shi, 2006).

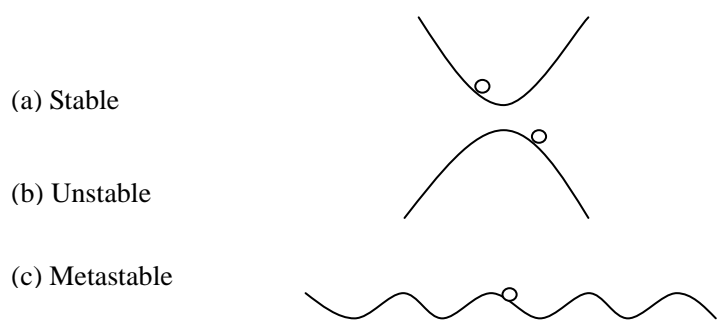

(a) Stable

(b) Unstable

(c) Metastable

Figure 3 Multi-possible equilibrium points in a metastable state

It is very interesting that there are multi-possible equilibrium points in a metastable state (see Figure 3). Stationary equilibrium is metastable in stock market (Shi, 2006). It specifies the dynamic equilibrium that trading price goes upward and downward in reference to an outcome, an agreement price, or a stationary equilibrium price in a time interval which generates a mean return of zero or very close to zero, following a small supply and demand imbalance; but the equilibrium price jumps and results in a significant price mean return from time to time after a larger supply and demand imbalance (Shi, 2006). A stationary equilibrium price is the price on a trading day at which the corresponding volume is the maximal (see Figure 1). It is close to a price mean value.

If trading price is higher than a stationary equilibrium price, then selling quantity will increase and buying quantity will decrease. Trading price will drop. On the other hand, if trading price is lower than the price, then selling quantity will decrease and buying quantity will increase. Trading price will rise. As a result, total trading volume gradually shows a certain distribution pattern over a price range and achieves a maximal at the price (see Figure 1).

In addition, stationary equilibrium might occur because of a limited resource of liquidity. If trading price deviates and is higher than a stationary equilibrium price, then market crowd could buy fewer shares with the same amount of money. The demand quantity reduces. Trading price will drop. If trading price deviates and is lower than a stationary equilibrium price, then market crowd can buy more shares with the same amount of money. The demand quantity increases. Trading price will rise. In a word, market crowd are forced to reach an agreement on a stationary equilibrium price, constrained by trading budget or wealth.

According to prospect theory (Kahneman and Tversky, 1979), status quo, social norms, and aspiration levels may determine a stock value. To meet trader's own satisfying feelings (Simon, 1956), a stock value shall include private information value (Kyle, 1985), speculative value (Hirshleifer, Subrahmanyam, and Titman, 2006; Han and Kumar, 2013), sentiment value (Barberis, Shleifer, and Vishny, 1998; Baker and Wurgler, 2006; Han, 2007), attention value (Da, Engelberg, and Gao, 2011), gamble value (Kumar, 2009), and entertainment value (Dorn and Sengmueller, 2009; Hoffmann and Shefrin, 2012) etc. So, asset prices might not equal to its fundamental value that is hardly calculated in stock market.

In summary, market crowd trade a stock in reference to an outcome or a stationary equilibrium price once they have made trading decision. They generate a reversal and negative feedback force in relative to an outcome or a stationary equilibrium price. As a result, trading price deviates from the equilibrium price and behaves like price momentum effect in a short term, but shows price reversal effect in a longer term and results in a mean return of zero or very close to zero on a trading day, following a small supply-demand imbalance. Market behaves stationary equilibrium widely on a trading day (Shi, 2006). It is about 94.34% in test (Shi, 2006).

So, we assume that market crowd trade a stock in simply heuristics (Tversky and Kahneman, 1974) and efficient adaptation, gradually tend to generate an outcome or a stationary equilibrium price in interaction and competition among themselves, and form a common belief (agreement) on a stock value (asset price) on a trading day.

### 3.3 Market Crowd's Adaptive Behavior to Gain and Loss

However, there is the other case that buying quantity does not reduce when price goes up. A stationary equilibrium state is metastable in stock market, similar to (c) in figure 3. When a large supply-demand imbalance breaks equilibrium, results in a stationary equilibrium price jump, and generates a significant price mean return from time to time, market crowd adapt to another stationary equilibrium price on a trading day. It is about 5.5% (34 over 618), a considerable number of events (Shi, 2006). According to behavior analysis (Skinner, 1938; Pierce and Cheney, 2004; Staddon, 2010), market crowd adapt to gain and loss or jump in a stationary equilibrium price by trading frequency. In general, they trade more frequently if they are previously rewarded by market gain and trade less frequently if they are previously punished by market loss[20].

|  |  | Hedonic Category | |
|---|---|---|---|
|  |  | Good (Gain) | Bad (Loss) |
| Preference in Choice | Buy | Volume Increase | Volume Decrease |
|  | Sell | Volume Increase | Volume Decrease |

Figure 4: Trading behaviors approaching good and avoiding bad

According to statistics and probability in mathematics, the frequency of the trading action is close to the probability of trading volume if any a trading volume is much great less than total trading volume on a trading day. So, we assume that market crowd adapt to gain and loss by trading volume increase or decrease significantly in interaction with market environment in any two consecutive trading days (see Figure 4). We will test two kinds of adaptive trading behaviors next.

## 4. EMPIRICAL TESTS

We test two kinds of market crowd's trading behavioral hypotheses relevant to the volume uncertainty associated with price, using high frequency data over more than two years in China stock market. When we examine hypothesis one, we focus on adaptive trading behaviors to an outcome only because cognitive biases caused by judgment in heuristics in terms of information and news have already been documented (Kahneman and Tversky, 1973; Tversky and Kahneman 1974).

### 4.1 Data

We use tick by tick high frequency data in Huaxia SSE 50ETF (510050) in China stock market from April 2, 2007 to April 10, 2009, a more than two year term when it experiences a whole course from bubble growth, burst, and shrink to reversal again. There are about 740 days and 495 trading days in more two years, in which stock prices are highly overestimated at one time but underestimated at another time. Thus, there are 495 volume distributions over a price range in our tests. The data is from HF2 database in Harvest Fund Management Co., Ltd.

---

[20] Investor's trading behavior with expectation on return is much more intelligent and complex than animal's operant behavior for foods. People are better adaptive to change in environment. For example, they might take buy and hold strategy (trade less frequently) to expect more gain against gain (see Section 4.3).

We process the data in two steps. First, we reserve two places of decimals in price by rounding-off method and add volume at a corresponding price[21]. Second, trading volume at a price is divided by total volume in a trading day. Thus, we obtain volume probability at a price and a volume distribution over a trading price range.

### 4.2 Models and Test Results

We test hypothesis one by regression model(s) and hypothesis two by correlation analysis, respectively.

4.2.1 Tests for hypothesis one (an estimated stock value hypothesis)

We test an estimated stock value hypothesis using two sets of volume distribution eigenfunctions over a price range. They are market crowd's optimal trading volume functions across price. They are analytical and closed form solutions from Shi's price-volume differential equation (Shi, 2006).

We run test by regression model(s). One set of the functions are:

$$|\psi_m(p)| = C_m |J_0[\omega_m(p - p_0)]|, \qquad (m = 0,1,2,\cdots) \qquad (1)$$

where $J_0$ is zero-order Bessel eigenfunction; $p_0$ and $p$ is a stationary equilibrium price and trading price, respectively; $\omega_m$ is an eigenvalue and angular frequency; $|\psi_m(p)|$ is a trading volume probability, the frequency of market crowd's trading action, and the intensity of their trading preference at a price $p$ over a price range on a trading day; and $C_m$ is a normalized constant (see Figure 5). The function describes market crowd's agreement trading behavior in interaction among them in stock market.

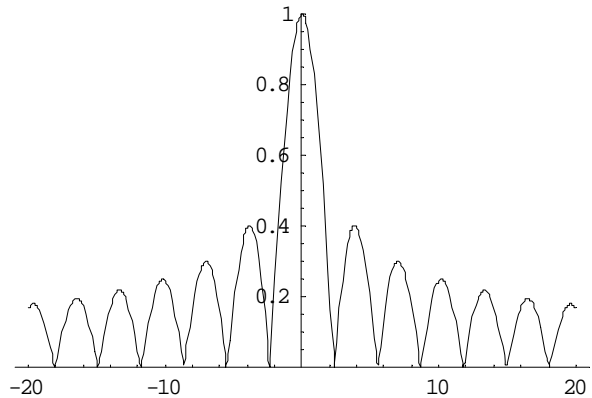

Figure 5: The absolute of zero-order Bessel eigenfunctions[22]

The other set of the functions are:

---
[21] Original data reserves three places of decimals.
[22] In Figure 4, x-coordinate is price and y-coordinate is trading volume probability in a time interval. The origin is a stationary equilibrium price.

$$\left|\psi_{n,m}(p)\right| = C_{n,m} e^{-\sqrt{A_{n,m}}|p-p_0|} \cdot \left|F\left(-n,1,2\sqrt{A_{n,m}}|p-p_0|\right)\right|, (n,m = 0,1,2,\cdots) \quad (2)$$

where $F\left(-n,1,2\sqrt{A_{n,m}}|p-p_0|\right)$ is a set of multi-order eigenfunctions, $n$ is the order of the function, and $A_{n,m}$ is an eigenvalue constant (see Figure 6). The function describes market crowd's independent trading behavior in stock market.

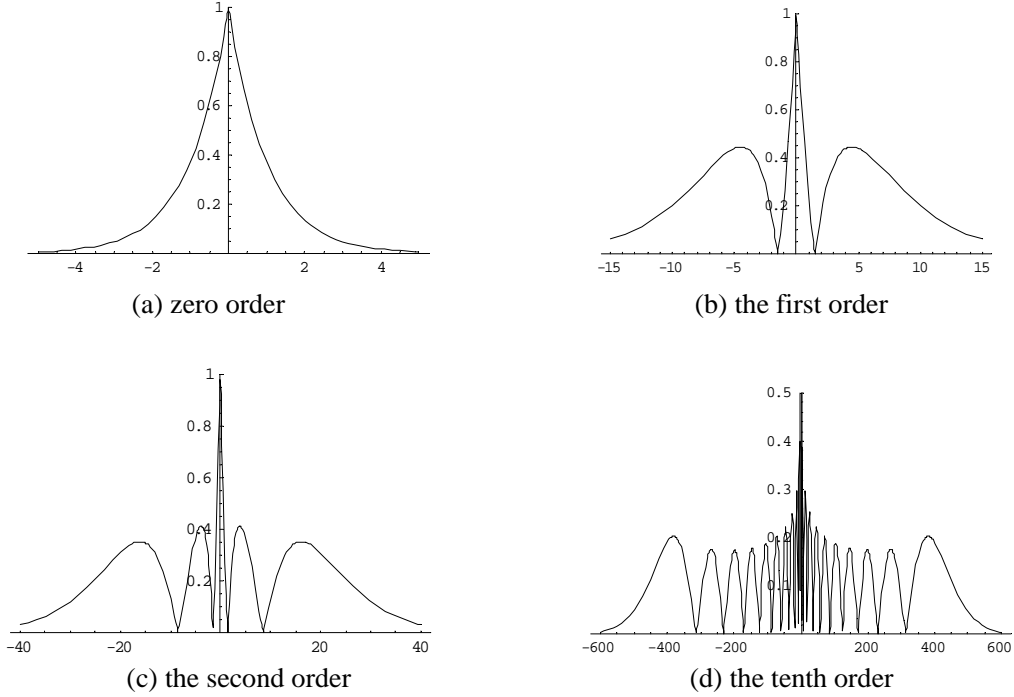

(a) zero order  (b) the first order
(c) the second order  (d) the tenth order

Figure 6: The absolute of the multi-order eigenfunctions[23]

We test a stationary equilibrium price or market crowd's agreement on a price using equation (1) because of several reasons. First, it captures market crowd's interactive and competitive behavior in stock market, which is ignored by previous models such as an economic Brownian motion model (Samuelson, 1965), an option pricing model (Black and Scholes, 1973), and an intertemporal CAPM (Merton, 1973). Second, it is consistent with a vast number of findings that there are high peaked, heavy tailed, and clustered characteristics in high frequency data test (Mandelbrot, 1963; Engle, 1982; Bollerslev, 1986; Mantegna and Stanley, 1995)[24]. Third, it could provide an explicit and quantitative explanation for volume distribution over a trading price range in a stationary equilibrium state. Fourth, it is gradually like an exponent distribution if its eigenvalue is larger. That is, it could cover market crowd's independent trading characteristics in which one of trading volume distributions is exponential. However, the major reason is that we attempt to explore a unified market theory by a price-volume probability differential equation (Shi, 2006) from which equations (1) and (2) are obtained (see Section 5).

---

[23] In Figure 5, x-coordinate is price and y-coordinate is trading volume probability in a time interval. The origin is a stationary equilibrium price.
[24] The volume distribution is close to return frequency distribution, according to the law of large number in probability and statistics in mathematics.

If test shows significance, then hypothesis one holds true. If test does not show significance, then we test it using a linear superposition model of equation (1). There is a stationary equilibrium price jump and a significant price mean return on a day if such a test shows significance. We test the rest that do not show significance, using the first-order eigenfunction in equation (2).

Our test results show that: 1) Tested by equation (1), 408 out of 495 distributions (about 82.42%) show significance. Market crowd trade a stock in efficient adaptation to an outcome, an agreement price, or a stationary equilibrium price (asset price) widely on a trading day; 2) Tested by a linear superposition model of equation (1), 59 out of 495 distributions (about 11.92%) show significance. There is jump in a stationary equilibrium price which results in a significant price mean return. Market crowd adapt to reach agreement on another stationary equilibrium price from time to time; 3) Tested by the first-order function of equation (2), 23 out of 495 distributions (about 4.65%) show significance. Market crowd tend to behave considerably no agreement on any a price; and 4) 5 remainders (about 1.01%) behave uniform distribution. Market crowd behave no agreement on any a price at all. In TableⅠare test reports, consistent with the prior findings by Shi (2006). Please read Supplement B for test details.

TableⅠ: Test Reports on Stationary Equilibrium and Agreement Price(s)

|  | No. of Distributions | Percentage (%) |
| --- | --- | --- |
| Total Number of Distributions | 495 | 100 |
| Agreement and Stationary Equilibrium | 408 | 82.42 |
| Agreement on Two Prices (Jump and Adaptive Behavior) | 59 | 11.92 |
| Agreement on Three Prices (No Agreement) | 23 | 4.65 |
| Uniform Agreement (No Agreement) | 5 | 1.01 |

Note: It is 4 trading hours per day in China stock market.

We have robust test in several terms in more than two years, in which Shanghai Securities Exchange Composite Index experiences from 3252.59 points to 6124.04 points, then drops to 1664.04 points, and reverse to 2444.25 points at the end. From the test reports above, we conclude that market crowd trade a stock not only in heuristics in terms of information and news (Kahneman and Tversky, 1973; Tversky and Kahneman, 1974), but also in adaptation to an outcome, an agreement price, or a stationary equilibrium price. They adapt to trade a stock in terms of an outcome rather than a fundamental value and form belief on a stock value in interaction and competition among themselves on a trading day. It holds true no matter whether the asset price(s) is obviously underestimated or overestimated. The trading action in terms of an outcome is quite useful for survival in natural selection but often leads cognitive biases and mispricings on a stock value, which is either overestimated or underestimated. Obviously, adaptation to a fundamental value in neoclassical paradigm is a special case of adaptation to an outcome or an asset price in behavioral framework where the outcome is equal to the fundamental value.

4.2.2 Test for hypothesis two (adaptive trading hypothesis)

We test an adaptive trading hypothesis by correlation analysis between a stationary equilibrium price jump (price mean return) and change in total trading volume in any two consecutive trading days. The hypothesis holds true if there is significant correlation between them (both positive and negative), because market crowd reach agreement on a stationary equilibrium price widely on a trading day.

We measure a price mean return $\bar{r}$ by stationary equilibrium price jump in any two consecutive trading days as

$$\bar{r} = \frac{\Delta p}{p_0}, \quad (3)$$

where $p_0$ is an equilibrium price on the first trading day and $\Delta p$ is an equilibrium price jump between two days.

We measure total trading volume increase or decrease $\Delta V$, which is corresponding to a jump in a stationary equilibrium price, in any two consecutive trading days by

$$\Delta V = \frac{V' - V}{V}, \quad (4)$$

where $V$ and $V'$ are total volume on T and T+1 trading days, respectively; Obviously, $\Delta V$ could be positive, negative, or zero.

We make correlation analysis between a stationary equilibrium price jump which is expressed in terms of equation (3) and change in total trading volume in any two consecutive trading days which is expressed in terms of equation (4).

We subdivide a more than two year term into 5 terms in which there are bubble growth, burst, and shrink until market reversal again in China, a whole course that is paralleled with the growth, burst, and collapse of sub-prime bubble that originated in the United States in 2008 and set off a chain reaction worldwide (reference to Table Ⅱ). This will provide a close look how market gain and loss, the outcome of prior trading action, influences the frequency of market crowd's trading action in response to information and news over time and occasion.

The first term is from April 2, 2007 (SSE Composite Index at 3252.59 points) to June 29, 2007 (SSE Composite Index at 3820.70 points), the first half before bubble burst in China. The second is from July 2, 2007 (SSE Composite Index at 3836.29 points) to October 31, 2007 (SSE Composite Index at 5954.77 points), the second half before bubble burst in China. The third is from November 1, 2007 (SSE Composite Index at 5954.77 points) to April 40, 2008 (SSE Composite Index at 3693.11 points), the first half after bubble burst in China. The forth is from May 5, 2008 (SSE Composite Index at 3761.01 points) to October 31, 2008 (SSE Composite Index at 1728.79 points), the second half after bubble burst in China. And the last is from November 3, 2008 (SSE Composite Index at 1719.77 points) to April 10, 2009 (SSE Composite Index at 2444.23 points), price reversal time interval after one year deep drop (reference to TableⅡ).

Table Ⅱ: Test Reports on Correlation and Its Significance

|   | Terms | Number of Distributions | SSE Composite Index (1A0001) | Correlation Coefficients and Its Significant Test Results |
|---|---|---|---|---|
| A | 2007.4.2—2009.4.10 | 494 | 3252.59—2444.23 | 0.1391 (t=3.115>$t_{crit}$=1.960) |
| B | 2007.4.2—2007.6.29 | 59 | 3252.59—3820.70 | -0.2567 (t=2.006> $t_{crit}$ =2.001) |
| C | 2007.7.2—2007.10.30 | 83 | 3836.29—5954.77 | **0.0729** (**t=0.6583< $t_{crit}$ =1.990**) |
| D | 2007.11.1—2008.4.30 | 122 | 5914.28—3693.11 | **0.1026** (**t=1.130< $t_{crit}$ =1.980**) |
| E | 2008.5.5—2008.10.31 | 123 | 3761.01—1728.79 | 0.1963 (t=2.202> $t_{crit}$ =1.980) |
| F | 2008.11.3—2009.4.10 | 107 | 1719.77—2444.23 | 0.4766 (t=5.556> $t_{crit}$ =1.983) |

Notes:
1) It specifies correlation between return and total volume probability increase or decrease in any two consecutive trading days;
2) Here, $t_{crit}$ is $t_{0.05/2}(n-2)$ ; If t> $t_{crit}$, then, correlation coefficient is significantly not equal to zero; otherwise, we can not reject original hypothesis that correlation coefficient is equal to zero;
3) It is printed in bold if test result does not show significance;
4) SSE Composite Index is measured by closing point.

We run test using Eviews 6.0. In TableⅡ are our test reports (please read Supplement B for details). Next, we are going to discuss test results.

### 4.3 Discussions on Test Results in TableⅡ

Investor's trading behavior with expectation on return is much more intelligent and complex than animal's operant behavior for foods. People prefer to trade more frequently in general if they have been previously rewarded by gain, but sometime they might trade less frequently for more gain. In TableⅡ, we find that there are significantly both positive and negative correlations between return and change in volume (reference to lines A, B, E, and F in TableⅡ). Market crowd are sensitive to gain and loss and adapt to them by trading volume increase or decrease significantly[25].

---

[25] According to behavior analysis, there must really be some general correlation between nonhedonic behaviors, the frequency of market crowd trading action which is represented by trading volume probability, and its hedonic consequences, gain and loss, in stock market (see Figure 4).

4.3.1 Positive correlation reveals not only disposition and reversal effects but also herd and momentum effects

Disposition is a kind of selling behavior. Shefrin and Statman (1985) term the disposition effect that investors have a desire to realize gains by selling stocks that have appreciated, but to delay the realization of losses. It is explained by prospect theory (Tversky and Kahneman, 1992) and consistent with the tests by Odean (1998a), Weber and Camerer (1998), Grinblatt and Keloharju (2001), Grinblatt and Han (2005), and Dhar and Zhu (2006) etc. Disposition sellers sell more in the face of gain or positive return and sell less in the face of loss or negative return. Obviously, they expect price reversal and their trading action generates price reversal effect.

Herd behavior exists widely in our social activities (Banerjee, 1992; Lux, 1995; Shiller, 1995). It has been theoretically linked to many economic activities. It is often said to occur when many people take the same action (Graham, 1999). There are many herd behaviors in stock market (Hirshleifer and Teoh, 2009). Herd buyers prefer to buy more when they observe that others make money when a significant positive mean return happens. They buy less when they watch that others loss money when a significant negative return occurs. It is a kind of observational, imitative, and social learning (Bandura, 1986). Herd buyers expect price movement continuation. Their trading action produces price momentum effect.

Therefore, disposition and price reversal effects coexist with herd and price momentum effects if there is a significantly positive correlation between return and change in volume (see Table Ⅲ). Such behavioral effects can be seen in our test results at lines A, E, and F in Table Ⅱ. It is interpreted in Figure 4. It is not inconsistent with the findings by Grinblatt and Han (2005) that investors with disposition effect cause price momentum in stock market.

Table Ⅲ. Trading Behaviors Matching with Correlations

| Types of Correlation | Trading Behaviors |
| --- | --- |
| Positive Correlation | disposition, herd, momentum, and reversal |
| The Strongest Positive Corr. | sustainable liquidity utility injection for market up reversal |
| Negative Correlation | greed, fear, caution, and brave |
| Insignificant Correlation | dysfunction between judgment and trading |

4.3.2 The strongest positive correlation makes known a necessary condition for market reversal at bottom

There is most significant positive correlation between return and change in volume when market reverses right after it experiences successive and steep drop in a year (reference to line F in Table Ⅱ). When SSE Composite Index drops from 6124.04 points at top to 1664.04 points at bottom in a year around, market crowd have already learned from previously severe punishment. They have behaved momentum and the strongest expectation on market punishment. Therefore, there is the most pronounced disposition effect to realize gains in short term. We can infer from this finding that price reversal takes place right after market crowd experience bearish if and only if participants make a larger sustainable liquidity utility[26] or energy injection to pair a very large volume of shares that are sold by disposition sellers (see Table Ⅲ).

4.3.3 Negative correlation shows greed and fear as well as caution and brave

Now, we explain a special case in which there is a significant negative correlation between return and change in volume in a time interval at line B in Table Ⅱ, although it could be observed in technical analysis from time to time in stock market.

Learning is a circular process in which surprise (the violation of expectation) and novelty (no expectation) cause one to update his representation of the situation, which in turn leads to new activity, hence to a new situation, perhaps more surprise, and so on in a spiral that usually converges on an adaptive pattern (Staddon, 2010). When SSE Composite Index kept steady rising from 998.23 points in July, 2005 to 3183.98 points in March, 2007 (the beginning of samples), market crowd had already been better rewarded if they took a buy-and-hold trading strategy. They naturally consider a context of similar outcomes for comparison, i.e. they consider more gains against gains or more losses against losses (McGraw et al., 2010). Stock holders take predictable price momentum strategy and sell less in adaptation when price goes up. They fear for worse and can't help to sell more in panic when severely punished in surprise by abrupt and deep drop. Meanwhile, cash holders buy less with expectation on price reversal when price mean return is positive, but buy more when the mean return is negative. They caution to buy when price goes up but brave to buy when it drops down (see Table Ⅲ). It is consistent with the finding by Grinblatt and Han (2005). The effect of disposition selling and herd buying disappears (see Figure 7).

|  |  | Hedonic Category | |
|---|---|---|---|
|  |  | Good (Gain) Worse (More Loss) | Better (More Gain) Bad (Loss) |
| Preference in Choice | Buy | Volume Increase | Volume Decrease |
|  | Sell | Volume Increase | Volume Decrease |

Figure 7: Adaptive learning comparing more gains (more losses) with gains (losses)

4.3.4 Insignificant correlation suggests dysfunctional behaviors

Finally, positive correlations lack significance right before and just after bubble burst in 2007 in China, respectively (reference to lines C and D in Table Ⅱ).

---
[26] Liquidity utility is expressed in terms of trading wealth. It is the rate of liquidity. Read details in Section 5.1.1

Disposition effect and herd behavior are not significant. On one hand, market crowd had desire to lock in gain and keep cash in safe when they ware aware of high risk because market index was very high right before bubble burst. On the other hand, they greed to ride on bubble for "maximum" return when they had been consistently rewarded by gain and a rising SSE Composite Index. It is behavioral dysfunction between judgment and trading.

Similarly, there is also behavioral dysfunction between judgment and trading just after the bubble burst. When market crowd are aroused, they behave high trading momentum which maintains their trading action in the face of risk. Although they are punished by SSE Composite Index drop, there is tremendous resistance to stop trading in stock market (see Table Ⅲ).

## 4.4 Summaries

We test two kinds of market crowd's adaptive trading hypotheses: market crowd's trading action in adaptive to an outcome or a stationary equilibrium price (with a mean return of zero or very close to zero), and market crowd's trading action in adaptive to gain and loss (a significant mean return) by trading volume increase or decrease. We conclude that: (1) Market crowd trade a stock in efficient adaptation, gradually tend to achieve agreement on an outcome or an asset price widely on a trading day, generate such a stationary equilibrium price in interaction and competition among themselves very often no matter whether the asset price is highly overestimated or underestimated; (2) Market crowd adapt to gain and loss by trading volume increase or decrease efficiently in interaction with market environment in any two consecutive trading days because there exists an agreement price or a stationary equilibrium price widely on a trading day. The adaptive trading behaviors, though useful for survival in natural selection, often lead cognitive biases and mispricings; and (3) Considering more gains (more losses) against gains (losses), market crowd adapt to trade less for more return when they are consistently rewarded in bull market (market crowd can't help to trade more in fear of more losses when they are severely punished in surprise by abrupt and large price drop). While the effect of herd buyers and disposition sellers disappears, other trading behavioral anomalies such as greed and fear might pronounce significantly. Market crowd trading behavior "anomalies" vary in response to gains and losses by trading volume increase or decrease over time and occasion.

According to prospect theory (Kahneman and Tversky, 1979), judgment in heuristics in terms of information and news often leads cognitive biases on stock value. In addition, we test and find the other cause of cognitive biases—judgment in adaptation to an outcome or an agreement price. It is obvious that adaptation to a fundamental value in neoclassical paradigm is a special case of adaptation to an outcome or an asset price in behavioral framework where the outcome is equal to the fundamental value. Heuristics and adaptation are insensitive each other (Kahneman and Tversky, 1973). Both together cause cognitive biases and result in long time mispricings between asset prices and fundamental values in stock market. We will consider them in the trading action to build a unified theory in Section 5.

## 5. AN EQUATION FOR MARKET CROWD'S TRADING BEHAVIORS

So far, we have tested two market crowd's trading behavioral hypotheses relevant to the volume uncertainty with price. They holds true in stock market. However, it is not the end but the beginning of our research. We are still much closer to the beginning of the research agenda than we are to the end, although there is a lot of accomplishment in a short period of time (Barberis and Thaler, 2003).

Now, we attempt to reconcile market dynamics and crowd's trading behaviors in a unified framework by deriving a price-volume differential equation (Shi, 2006).

### 5.1 A Liquidity Utility Hypothesis

Influenced by prospect theory (Kahneman and Tversky, 1979), Barberis, Huang, and Santos (2001) study asset prices in the assumption that investors derive a direct utility not only from consumption but also from change in the value of their financial wealth. It might help us to understand the liquidity utility hypothesis established from supply and demand law, behavior analysis, and prospect theory.

5.1.1 Hypothesis Three—the liquidity utility hypothesis

When we study a complex system such as stock market, we may consider it as a "black box" and study relationship between input and output regardless of what really happens inside[27]. This methodology might result in an unexpected and surprising outcome from time to time in research. Here, input is liquidity or total trading wealth, and output is the trading volume distribution over a price range on a trading day. We study how trading wealth or liquidity (input) constrains price (output) and trading volume (output) or how market crowd's trading action (input) generates a trading volume distribution over a price range (output), regardless of an individual's trading action that might change actual price path at a moment in "black box". The "black box" facilitates descriptive explanation for Shi's price-volume probability differential equation (Shi, 2006).

Constrained by budget, asset, and wealth, market crowd trade to produce liquidity and result in the rate of liquidity. The trading wealth at a price, $M(p)$, is equal to the product of price $p$ and total trading volume $v$ at this price, i.e.,

$$M = pv. \tag{5}$$

We define total trading volume $v$ at a price in a given time interval [0, t] as trading momentum (Shi, 2006)[28]. It is

$$v_t = \frac{v}{t}. \qquad (6)$$

---

[27] http://en.wikipedia.org/wiki/Black_box

[28] Total trading volume gradually emerges a pattern of distribution over a price range when it takes a longer time interval on a daily basis. It is similar to interference behavior in a beam of electrons passing through a double-slit in physics. Accumulative trading volume at a price in a time interval is the trading momentum. It is derived from action and defined critically in econophysics (Shi, 2006). Its dimension is [share][time]$^{-1}$.

Now, we assume the existence of a liquidity utility $U(p,v_{tt})$ expressed in terms of trading wealth[29]. It is the rate of liquidity, similar to power or the rate of work in physics, which we term liquidity energy $E(p,v_{tt})$ (Shi, 2006 and 2013). It is defined as follows:

$$U(p,v_{tt}) \equiv E(p,v_{tt}) = p \cdot v_{tt}, \qquad (7)$$

and

$$v_{tt} = \frac{v}{t^2} = \frac{v_t}{t}, \qquad (8)$$

where $v_{tt}$ is trading momentum force or impulse (Shi, 2006)[30].

A liquidity utility is a momentum utility because it is generated by trading momentum force or price momentum force $v_{tt}$ (see equations (7) and (25), and read Section 5.3.2).

As we discuss in Section 3.2, market crowd tend to sell (buy) more when price goes up (down). If everyone would maximize a liquidity utility when he trade a stock, then everyone would buy at the lowest price and sell at the highest price in a given time interval. Obviously, it is totally impossible unless there is no price change at all or there is an exclusive and monopoly trading price, i.e., a static state.

While some traders prefer to buy a stock, others prefer to sell at the same price. They trade a stock in a preference structure over an acceptable trading price range to meet their individual's satisfying feelings with time and resource constraint, not just at an exclusive price (Simon, 1955; Kahneman and Tversky, 1979). Whereas one waits for a while to buy (sell) a stock at a lower (higher) price in uncertainty, he also worries that he has to buy (sell) at a higher (lower) price later because of free competition from others. Market crowd trade a stock in simple heuristics and efficient adaptation over a price range on a trading day. There are a limited number of trading volume distributions over a price range in stationary equilibrium in stock market.

We classify market crowd into three groups of boundedly rational agents: momentum traders, reversal traders, and interactive traders who are influenced by both momentum and reversal traders[31]. The trading momentum $v_t$ pairs the same number of shares traded by reversal and interactive traders. That is, total number of shares traded in any a trading volume $v$ is $2v$. Moreover, the trading action that produces liquidity refers to not only the information and news—heuristics and representativeness in judgment, but also the outcomes of prior trading—adaptation by return reinforcement or punishment (see Figure 2). Heuristics and adaptation are insensitive each other (Kahneman and Tversky, 1973; Tversky and Kahneman, 1974). Both together often lead cognitive biases in decision making. Therefore, a liquidity utility or momentum utility could be divided into two parts by template learning and adaptive learning: a trading volume probability weighting utility relevant to judgment by heuristics and a reversal utility relevant to judgment by adaptation. So, we write

$$U(p,v_{tt}) \equiv P \cdot U(p,v_{tt}) + (1-P)U(p,v_{tt}) = \frac{v}{V} \cdot U(p,v_{tt}) + A(p-p_0)^\alpha, \quad (9)$$

where $p$ is trading price; $p_0$ is an agreement price, a stationary equilibrium price, or an asset price; $A$ is a coefficient; $\alpha$ is a constant which will be determined in Section 5.2.1; $P$ is trading volume probability which is equal to a total trading volume at a price $v(p)$ over a total trading volume across a trading price range $V$, that is,

$$P = \frac{v}{V}. \quad (10)$$

---

[29] The dimension of liquidity utility is [currency unit][time]$^{-2}$.

[30] Momentum force is derived from action or trading momentum (see equations (24) and (25) in section 5.3). Its dimension is [share][time]$^{-2}$. It is equivalent to the price momentum force that drives trading price deviation from a stationary equilibrium. The $v_t$ and $v_{tt}$ do not mean $\partial v/\partial t$ and $\partial^2 v/\partial t^2$, respectively (Shi, 2006).

[31] Although some traders are of intended rationality in a group, others might be boundedly rational or irrational. Thus, a group of them behave boundedly rational in general. Moreover, rational behavior is a special case of boundedly rational behavior when the outcome of crowd's trading action is equal to a stock fundamental value.

A trading volume probability weighting utility in equation (9) resembles and is close to an expected liquidity utility—the expected value of future trading wealth.

We consider transformations in phase space; that is, we shall assume that trading momentum $v_t$ is included in the transformations. Furthermore, we shall restrict ourselves to canonical transformations which preserve the Hamiltonian form of equation in the new variables (Greenwood, 1977).

In stationary equilibrium, we may separate the liquidity momentum utility $U(p,v_{tt})$ by trading momentum $v_t$ and divide it into two parts in terms of canonical variables—price $p$ and trading momentum $v_t$. It is

$$U(p,v_{tt}) = I(p,v_t) + W(p-p_0) = I(p,v_t) + A(p-p_0)^\alpha, \quad (11)$$

where $W(p-p_0)$ is reversal utility or reversal energy that restores trading price to a stationary equilibrium price $p_0$ as they occur in the market[32], the rest $I(p,v_t)$ is an interactive or competitive utility which is relevant to interaction or competition between momentum traders and reversal traders (see equations (12), (17) and (38). Compared equation (9) with equation (11), we have interactive or competitive utility

$$I(p,v_t) = P \cdot U(p,v_{tt}) = \frac{v}{V} \cdot p \cdot v_{tt} = \frac{v}{V} \cdot p \cdot \frac{v}{t^2} = \frac{p}{V}\left(\frac{v}{t}\right)^2 = p\frac{v_t^2}{V}. \quad (12)$$

We recast equation (11) using equations (7) and (12) and have

$$-E + p\frac{v_t^2}{V} + W(p-p_0) = 0. \quad (13)$$

**Hypothesis Three (A Liquidity Utility Hypothesis):** A liquidity or momentum utility is equal to the sum of an interactive utility and a reversal utility.

Equation (13) is a holonomic constraint condition on price $p$ and trading momentum $v_t$. Equation (13) is effective because equation (9) always holds true in stock market. By equation (13), we will derive a price-volume probability differential equation in Section 5.3.

5.1.2 Hamiltonian in finance

---

[32] In a stationary equilibrium state, the potential could be viewed as time and momentum independent if the reversal force is a constant. See equation (15).

We define that the sum of interactive or competitive utility $I(p,v_t)$ and reversal utility $W(p-p_0)$ in equation (13) is Hamiltonian $H(p,v_t)$ in finance. That is,

$$H(p,v_t) = I + W = p\frac{v_t^2}{V} + W(p - p_0). \tag{14}$$

The interactive or competitive utility $I(p,v_t)$ is similar to an expected utility—the expected value of future trading wealth. The reversal utility $W(p-p_0)$ in reference to an outcome or a stationary equilibrium price $p_0$ is not considered in utility theory in neoclassical finance. We will find out optimal trading momentum function(s) by solving Shi's price-volume probability differential equation (see Sections 5.3 and 5.4).

### 5.2 Stationary Equilibrium and Its Mathematical Expressions

We have documented that stationary equilibrium exists widely in stock market (see Sections 3.2, 4.1, and 5.1). It can be expressed by a reversal utility, $W(p-p_0)$, which is a function of price in reference to an outcome, an agreement price, or a stationary equilibrium price.

#### 5.2.1 Stationary equilibrium and its measurement

The reversal utility $W(p-p_0)$ in equation (13) describes trading price behavior following a small supply-demand imbalance. It restores trading price toward a stationary equilibrium price and results in a mean return of zero or very close to zero on a trading day. The dimension of the reversal utility is the same as that of the momentum utility. Thus, the reversal utility is proportional to price. It is a linear utility, i.e. $\alpha = 1$ in equation (9) or (11). It can be expressed in mathematical language by

$$W(p - p_0) = A(p - p_0) \approx A(p - \bar{p}), \tag{15}$$

where $p_0$ is an outcome, an agreement price, or a stationary equilibrium price; and $\bar{p}$ is a volume probability weighted price mean value. $A$ is a coefficient which will be discussed next. It could be a constant or a variable (See Section 5.3).

According to econophysics (Shi, 2006), we have price reversal force $F_r$ in stationary equilibrium using equation (15) as

$$F_r = -\frac{\partial W}{\partial p} = -A, \tag{16}$$

where $-A$ is reversal force[33] in which the minus sign indicates that the direction of the force is always toward a stationary equilibrium price $p_0$. Reversal force is a necessary and sufficient condition for stationary equilibrium. It could not be equal to a constant over a price range. We will discuss it in Section 5.3.

Differentiating equation (13) by using equations (7) and (15), we have

$$v_{tt} - A - \frac{v}{V}v_{tt} = 0, \tag{17}$$

where $v_{tt}$ is the price momentum force (see equation (25) in Section 5.3), $-\frac{v}{V}v_{tt}$ is interactive or competitive force in which the minus sign indicates that the direction of the force is always toward an outcome or a stationary equilibrium price. In stationary equilibrium, the sum of momentum force, reversal force, and interactive force is equal to zero.

---

[33] The dimension of the reversal force is [share][time]$^{-2}$.

5.2.2 Distinction between stationary equilibrium and traditional equilibrium

Contrary to "law of one price" in traditional equilibrium in economics, a stationary equilibrium price is not exclusive one at a time for a stock. As long as larger supply and demand imbalance occurs, for example, institutional speculative trading action, it could result in a jump in stationary equilibrium price even although there is no change in its fundamental value at all. In addition, stationary equilibrium is generated by trading action—adaptation in terms of the outcome of trading action and heuristics in terms of information and news. There is a shift of an outcome (Kahneman and Tversky, 1979), and there is a jump in a stationary equilibrium price from time to time. Thus, there is no one-to-one corresponding relation between a stationary equilibrium price and the fundamental value that is hardly calculated in stock market. A stationary equilibrium price includes not only a fundamental value but also a variety of behavioral values, thus allowing a long time and larger mispricing between the asset prices of a stock and its fundamental value.

### 5.3 The Price-volume Probability Differential Equation

Now, it is ready for us to have Shi's differential equation from a liquidity utility hypothesis.

5.3.1 The price volume probability differential equation

Suppose that market behaves a probability wave in price and volume coordinates (Shi, 2006). Its trading volume function $\psi(p)$ satisfies

$$\psi(p) = \operatorname{Re}^{iS/B}, \tag{18}$$

where $R$ is the wave amplitude, $S$ its trading action or Hamiltonian principal function, $B$ a constant to make its phase dimensionless, and $i$ is an imaginary number which $i^2=-1$ (Schrŏdinger, 1928; Derbes, 1996; Zeng, 2000) [34].

We assume that market satisfies a Hamilton-Jacobi equation in a phase or momentum space (Greenwood, 1977), i.e.,

---

[34] If we assume $\psi(p) = \operatorname{Re}^{S/B}$, we have the same result (Schrŏdinger, 1928 and Derbes, 1996).

$$\frac{\partial S}{\partial t} + H\left(p, \frac{\partial S}{\partial p}\right) = 0, \tag{19}$$

where the Hamiltonian is

$$H\left(p, \frac{\partial S}{\partial p}\right) = I\left(p, \frac{\partial S}{\partial p}\right) + W(p - p_0), \tag{20}$$

and

$$Q \equiv \frac{\partial S}{\partial p}. \tag{21}$$

Equation (21) is a generalized momentum in price coordinate. Equations (13) and (19) are equivalent. The Hamilton-Jacobi partial differential equation (19) is expressed by an action or principal function in canonical transformations.

The assumption is acceptable in equation (19). If the price-volume differential equation we derive from equation (19) could describe real world in stock market, then the assumption holds true. Otherwise, it should be rejected.

To solve equation (19), we have

$$-\frac{\partial S}{\partial t} = H\left(p, \frac{\partial S}{\partial p}\right) = E. \tag{22}$$

Equation (22) holds true as long as $E$ is price and time independent no matter whether $E$ is a constant, because equation (13) is effective anyway. In addition, if $\frac{\Delta p}{p} \to 0$ and if $\frac{v}{V} \to 0$, then $\frac{I}{E} \to 0$ and $\frac{I}{W} \to 0$. Therefore, when we study the interactive utility $I$, both momentum utility $E$ and reversal utility $W$ may be approximately regarded as infinity or a constant.

From $-\frac{\partial S}{\partial t} = E$ in equation (22), we have a special solution[35]

$$S(p, v_t, t) = S_1(p, v_t) - Et = pv_t - Et. \tag{23}$$

In canonical transformations, we have an action or trading momentum Q from equation (21) as

$$Q \equiv \frac{\partial S}{\partial p} = v_t. \tag{24}$$

According to definition in classical dynamics (Greenwood, 1977), moreover, we have action force, impulse, or price momentum force $F_m$,

$$F_m \equiv \frac{Q}{t} = \frac{v_t}{t} = \frac{v}{t^2} = v_{tt}. \tag{25}$$

So, we rewrite equation (13) expressed by an action or principal function in canonical transformations, that is, $H\left(p, \dfrac{\partial S}{\partial p}\right) = E$ in equation (22), as

$$-E + \frac{p}{V}\left(\frac{\partial S}{\partial p}\right)^2 + W(p - p_0) = 0. \qquad (26)$$

---

[35] The dimension of the trading action is [currency unit][time]$^{-1}$. It is the same as that of liquidity.

From equation (18), we have

$$\frac{\partial S}{\partial p} = -\frac{iB}{\psi}\frac{\partial \psi}{\partial p}, \qquad (27)$$

then, we put equation (27) into equation (26), make due allowance for the complex nature of $\psi$, and construct a Lagrange function $L(p, \psi)$, that is,

$$L(p, \psi) = (W - E)\psi^* \psi + \frac{B^2}{V} p \left(\frac{d\psi^*}{dp}\right)\left(\frac{d\psi}{dp}\right). \qquad (28)$$

**Hypothesis Four (The Least Price or Cost Variation Principle):** Market crowd trade a stock in heuristics and adaptation to produce the least price or cost variation in competition among themselves at the moment[36].

In financial market, market crowd trade a stock in a preference structure over a price range, not just at an exclusive price, to meet their individual's satisfying feelings with time and resource constraint (Simon, 1956). They trade in both heuristics in terms of information and news, and adaptation to an outcome by reinforcement and punishment. It generates stationary equilibrium. Total trading volume gradually shows distribution over a price range on a trading day. However, it is an outstanding problem how trading price is actually volatile in "black box".

According to a trading rule in stock market, "price first and time first", one pairs bid or ask shares at the price that is most close to current trading price first. For example, if current trading price is $8.00, next trading priority is given to ask volume at price $8.01 instead of $8.02 unless total ask volume at $8.01 has already been paired. In another word, a seller who asks for $8.02 has to wait for a while until total ask volume at $8.01 has already been sold. Whereas one waits for a while to buy (sell) a stock at a lower (higher) price, he also worries that he has to buy (sell) at a higher (lower) price later because of free competition from others in stock market. He is interacted and competitive with other traders by bid and ask. He accepts current price, trades a stock in heuristics and adaptation, and produces the least price or cost variation in competition with others at the moment. Thus, it leads us to have the least trading price or cost variation principle.

If current trading price is $p_c$, then next trading priority is given to minimize price or cost variation with respect to it. It states that actual trading price path is determined by a liquidity utility functional to minimize its function $\psi(p)$ with respect to price or cost (price and time) variations. Its mathematical expression is

$$\delta \int L(p,\psi)dp = 0. \tag{29}$$

---

[36] Regarding to calculus of variations, please read at http://en.wikipedia.org/wiki/Calculus_of_variations

From the least trading price or cost variation principle—equations (29), Shi (2006) derives a time-independent price-volume probability differential equation,

$$\frac{B^2}{V}\left(p\frac{d^2\psi}{dp^2} + \frac{d\psi}{dp}\right) + [E - W(p - p_0)]\psi = 0. \tag{30}$$

Equation (30) describes market behavior in price and volume coordinates. It can be solved to obtain the extremal function that describes market crowd's optimal or preference choice over a price range in trading action. The abstract function, $|\psi(p)|$, is trading volume distribution function in a time interval, i.e., trading momentum and preference distribution. We will solve it in Section 5.4.

5.3.2 Trading momentum, price momentum, and preference

According to equation (24), trading momentum is $v_t$. It can be understood not only for the trading momentum that maintains market crowd to trade frequently in the face of risk but also for the price momentum that market crowd trade to drive price deviation from a stationary equilibrium price. The larger the trading momentum $v_t$ is, the larger the momentum force $v_{tt}$ is. The momentum force drives price deviation from a stationary equilibrium price, according to equation (25). Thus, trading momentum is equivalent to price momentum[37]. In addition, the momentum distribution we observed also represents market crowd's trading preference. They prefer to trade more volume at one price than another in interaction or competition among themselves, not trade randomly or uniformly. Market crowd trade in heuristics and adaptation, generated an outcome and stationary equilibrium in interaction or competition among themselves, and achieve an agreement on asset price(s) (see Section 4.4). They trade most at a stationary equilibrium price on a trading day.

5.4 A Market Crowd's Independent Trading Price Model and
A Market Crowd's Agreement Trading Price Model

Now, we are going to develop two typical kinds of behavioral models: a market crowd's independent trading price model and a market crowd's agreement trading price model, based on the price-volume probability wave equation (30).

Substituting equation (15) into equation (30), we have

$$\frac{B^2}{V}\left(p\frac{d^2\psi}{dp^2} + \frac{d\psi}{dp}\right) + [E - A(p - p_0)]\psi = 0. \tag{31}$$

Choosing natural unit $\frac{V}{B^2} = 1$ and natural boundary conditions,

---

[37] It is consistent with the finding in behavior analysis in which response strength is essentially equivalent to behavioral momentum—resistance to change (Staddon, 2010).

$$\psi(0)=0, \quad \psi(p_0)<\infty \quad \text{and} \quad \psi(+\infty)\to 0, \tag{32}$$

then we obtain two sets of analytical and closed form solutions (Shi, 2006).

5.4.1 A market crowd's independent trading price model

In equation (31), if *E=constant*, then, we have a set of multi-order eigenfunctions. They are as follows:

$$\psi_{n,m}(p) = C_{n,m} e^{-\sqrt{A_{n,m}}|p-p_0|} \cdot F\left(-n, 1, 2\sqrt{A_{n,m}}|p-p_0|\right), (n,m=0,1,2,\cdots) \tag{33}$$

and

$$\sqrt{A_{n,m}} = \frac{E_{n,m}}{1+2n} = const. > 0, \qquad (n,m=0,1,2,\cdots) \tag{34}$$

where $F\left(-n, 1, 2\sqrt{A_m}|p-p_0|\right)$ are a set of n-order confluent hypergeometric eigenfunctions (the first Kummer's eigenfunctions), $E_{n,m}$ is liquidity utility or energy, $A_{n,m}$ is the magnitude of reversal force, in which the direction of the reversal force is always toward an equilibrium price, and $C_{n,m}$ are a set of normalized constants.

According to equation (34), the magnitude of the reversal force is a constant and is direct proportional to the square of a constant liquidity utility over a trading price range.

The absolute of functions (33) are

$$|\psi_{n,m}(p)| = C_{n,m} e^{-\sqrt{A_{n,m}}|p-p_0|} \cdot \left|F\left(-n, 1, 2\sqrt{A_{n,m}}|p-p_0|\right)\right|, (n,m=0,1,2,\cdots) \tag{35}$$

where $|\psi_{n,m}(p)|$ is accumulative trading volume probability, the frequency of market crowd's trading action, and the intensity of their trading preference at a price in a time interval (see Figure 8).

If market crowd behaviors are governed by equation (35), then, market crowd are independent. For examples, if n=0, then, we have $F\left(0, 1, 2\sqrt{A_{0,m}}|p-p_0|\right)=1$. The intensity of market crowd's trading preference $|\psi_{0,m}(p)|$ follows an exponent and log-normal distribution (see (a) in Figure 8). The distribution is consistent with that of price independent Brownian motion in neoclassical finance (Samuelson, 1965). Market crowd trade independently. They are forced to trade most at a stationary equilibrium price because of strict liquidity resource constraint.

According to equation (34), we have $A=E^2_{0,m}=constant$ ($E_{0,m}>0$) in which both reversal force $A_{0,m}$ and liquidity utility $E_{0,m}$ are constants over a price range. If trading price deviates and is higher than a stationary equilibrium price, then, market crowd are forced to buy less volume because the liquidity utility expressed in terms of trading wealth is the same over a trading price range. The demand quantity is reduced. Trading price will drop. If trading price deviates and is lower than the stationary equilibrium price, then, market crowd are relax to buy more. The demand quantity is increased. Trading price will rise.

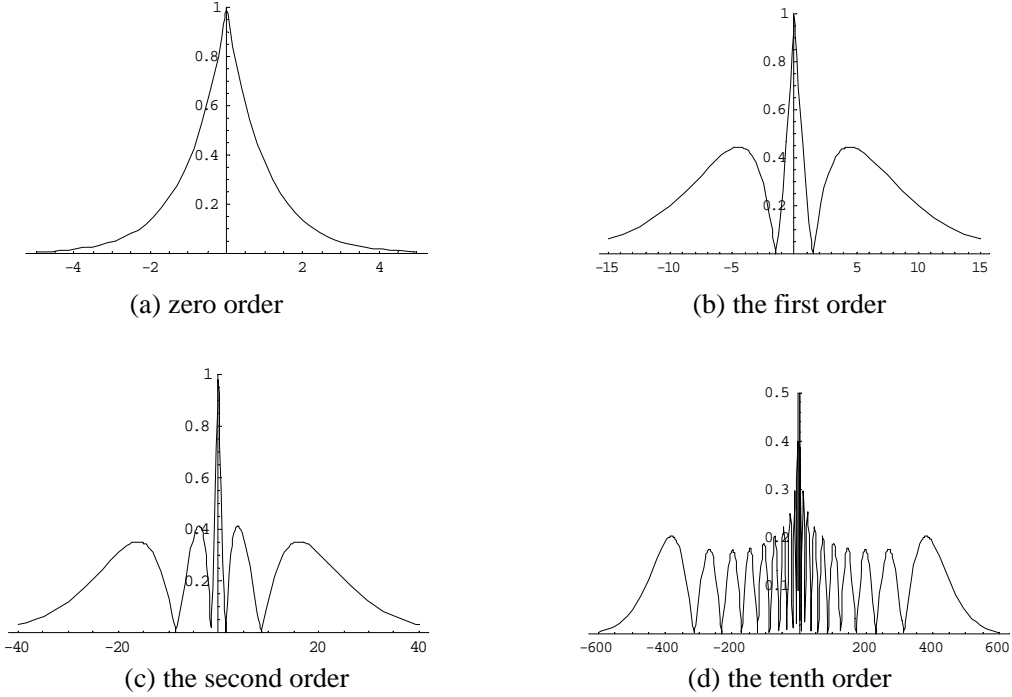

Figure 8: The absolute of the multi-order eigenfunctions[38]

If $n=1$, then, the intensity of market crowd's trading preference at a price in a time interval $|\psi_{1,m}(p)|$ follows the first order distribution and tend to be uniform distribution (see (b) in Figure 8). According to the reversal force $A_{1,m} = \frac{1}{9}E_{1,m}^2 = const.$, it is much weaker than $A_{0,m}$ if liquidity resource retains the same ($E_{1,m}=E_{0,m}$). Market crowd trade most at three prices and intend to have no agreement or preference to a price. Or market crowd trade most at three prices with much more liquidity resource ($E_{1,m}=9E_{0,m}$) if reversal force is the same $A_{1,m}=A_{0,m}$.

If $n=10$, then, the intensity of market crowd's trading preference in a time interval $|\psi_{10,m}(p)|$ follows the tenth order distribution (see (d) in Figure 8). The distribution is almost the same over a trading price range, i.e., there is no preference to a price. The magnitude of the reversal force is $A_{10,m} = \frac{1}{121}E_{10,m}^2 = const.$. It can be ignored in comparison with $A_{0,m}$ if $E_{10,m}=E_{0,m}$.

Specifically, if $n \to \infty$, then, the reversal force $A_{n,m} = \lim_{n \to \infty} \frac{E_{n,m}^2}{n^2} \to 0$. There are two possible solutions. First, market crowd trade in total freedom (reversal force is equal to zero) with limited liquidity resource. Second, market crowd trade with infinite liquidity resource ($E_{n,m}^2 = \lim_{n \to \infty} n^2 A_{n,m} \to \infty$) if the reversal force $A_{n,m} \neq 0$. It is not true in reality and is rejected[39]. Therefore, it is an acceptable solution that market crowd trade freely and independently with limited liquidity resource. For example, investors do not care short-term price volatility on a daily basis when they sell (buy) high overvalued (undervalued) stock for long term risk avoidance (value investment).

---

[38] In Figure 3, x-coordinate is price and y-coordinate is accumulative trading volume probability in a time interval. The origin is a volume weight price mean value.

According to equations (17) and (34), moreover, we have $A_{n,m}=constant$, $v_{(tt)n,m} \approx const.$ (if $\frac{\Delta p}{p} = \frac{p - p_0}{p} \to 0$), and $\left(\frac{v}{V} v_{(tt)n,m}\right)_{n,m} = v_{(tt)n,m} - A_{n,m} = const.$, that is, the sum of the three constant forces is equal to zero. It describes a static equilibrium in uncertainty, similar to static balance in mechanics and physics.

Therefore, if market crowd trade independently, then, 1) market crowd may be forced to trade most at a stationary equilibrium price because of a constant liquidity resource constraint over a price range; 2) market crowd may be forced to trade most at three prices and tend to have no agreement on a price because of more freedom and less liquidity resource constraint; and 3) market crowd may trade uniformly and have no agreement or preference to any a price at all. Market crowd's independent trading price model, equation (35), describes that the sum of three constant forces is equal to zero in stationary equilibrium, similar to static balance in mechanics and physics. It describes market crowd's independent and random trading behaviors.

5.4.2 A market crowd's agreement trading price model in interaction

In equation (31), if $E=p'v_{tt}$ ($p'=p-p_0$), then, we get the other set of analytical solutions. It is a set of zero-order Bessel eigenfunctions. They are written by

$$\psi_m(p) = C_m J_0[\omega_m(p - p_0)], \qquad (m = 0,1,2,\cdots) \qquad (36)$$

and

$$\omega_m^2 = v_{tt,m} - A_m, \qquad (\omega_m > 0) \qquad (m = 0,1,2,\cdots) \qquad (37)$$

where $J_0[\omega_m(p - p_0)]$ are a set of zero-order Bessel eigenfunctions, $\omega_m$ are a set of eigenvalue constants, $v_{tt,m}$ are a set of momentum force, $A_m$ are a set of reversal force, and $C_m$ are a set of normalized constants. According to equation (37), there is agreement force in interaction or competition when the sum of momentum force (a variable) and reversal force (the other variable) is equal to an eigenvalue constant over a trading price range.

From equations (17) and (37), we have

$$\omega_m^2 = v_{tt,m} - A_m = \frac{v}{V} v_{tt,m} = const., \qquad (\omega_m > 0; m = 0,1,2\cdots) \qquad (38)$$

---

[39] It is an underlying assumption in neoclassical finance that traders have unlimited borrowing and lending at the risk-free rate in CAPM and APT (Hirschey and Nofsinger, 2011). It is not true in reality and should be rejected.

where $\omega_m^2$ is angular frequency and $\dfrac{v}{V} v_{tt,m}$ is the magnitude of agreement force in interaction and competition. We measure agreement by a constant interactive or competitive force over a price range. It is $\omega_m^2 = \dfrac{v}{V} v_{tt,m} = const$.

The absolute of functions (36) is

$$|\psi_m(p)| = C_m |J_0[\omega_m(p - p_0)]|, \qquad (m = 0,1,2,\cdots) \qquad (39)$$

where $|\psi_{n,m}(p)|$ is the volume probability, the frequency of market crowd's trading action, and the intensity of their trading preference at a price in a time interval (see Figure 9).

The explicit model, equation (39), has open tails at two sides in reference to a stationary equilibrium price, allows infinite variance, and provides a long-neglected but important class of statistical distributions (see Figure 8)[40]. The volume distribution shows high peaked, heavy tailed, and attenuate clustered (damped cluster) characteristics, consistent with the findings by Mandelbrot (1963).

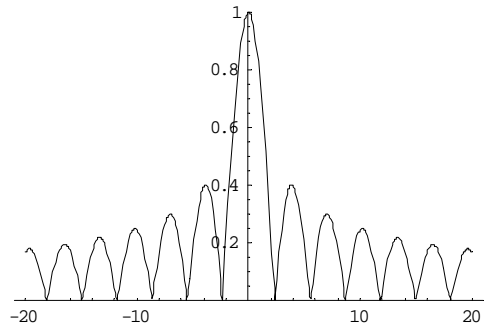

Figure 9: The absolute of zero-order Bessel eigenfunctions

There are several statistical characteristics between agreement trading distribution (see Figure 9) and independent trading distribution (see Figure 8). First, agreement trading distribution shows attenuate clusters over a trading price range in reference to a stationary equilibrium price, while independent trading distribution is almost uniform except for an exponential distribution. Second, agreement trading distribution has an open tail, whereas independent trading distribution is closed. Third, agreement trading distribution and exponential distribution are gradually alike if agreement force becomes larger. Fourth, the higher the order is in an independent trading distribution, the better it distributes uniformly. Fifth, both could show maximal volume at a stationary equilibrium price. Both might have a set of an infinite number of discrete extremal values (Please read Supplement A about eigenvalue and probability wave).

---

[40] Mandelbrot (1963) observes abnormal distribution in return and proposes a stable Paretian hypothesis allowing infinite variance, claiming that it is significantly different from independent and normal distribution. The finding ignites a long-term fundamental debate on price dynamic behavior in financial academic studies. Fama (1963) admits that Mandelbrot's stable Paretian hypothesis has focused attention on a long-neglected but important class of statistical distributions. The value of much work may be doubtful if Mandelbrot's hypothesis is upheld by data, since past research on speculative price has usually been based on statistical tools which assume the existence of a finite variance. Unfortunately, rigorous and analytical sampling theory will be difficult to develop as long as explicit expressions for the density functions are not known. The next step must be both to test the abnormal behavior on a broader range of speculative series and to develop more adequate statistical tools for dealing with the distributions.

### 5.5 A Market Crowd's Adaptive Trading Price Model

Now, let us explore dynamic mechanism in correlation analysis by equation (3) and equation (4), based on a price-volume probability differential equation (31).

We have tested in Section 4 that stationary equilibrium exists extensively in stock market, consistent with findings by Shi (2006) and Shi et al (2010). If a stationary equilibrium price is $p_0$, liquidity utility or energy is $E$, total trading volume is $V$, and restoring force is $A$ for a stock on a $T\text{-}1$ trading day, then we can assume that its market satisfies

$$\frac{B^2}{V}\left(p\frac{d^2\psi}{dp^2}+\frac{d\psi}{dp}\right)+[E-A(p-p_0)]\psi=0. \qquad (40)$$

If speculative institutions trade to cause a larger supply and demand imbalance and generate trading momentum force $v_{tt}'$ on $T$ trading day, then it is equivalent that we put a liquidity utility or energy increment $E'$ into Hamiltonian in equation (40). We write

$$\frac{B^2}{V}\left(p\frac{d^2\psi}{dp^2}+\frac{d\psi}{dp}\right)+[E-A(p-p_0)+E']\psi=0. \qquad (41)$$

Since differential equation (40) retains its validity on $T$ trading day, we can simplify equation (41) to be

$$\frac{B^2}{V_T}\left(p\frac{d^2\psi}{dp^2}+\frac{d\psi}{dp}\right)+[E_T-A_T(p-p_{T0})]\psi=0, \qquad (42)$$

where $p_{T0}=p_0+\Delta p$ is a stationary equilibrium price on $T$ trading day; $\Delta p$ is the stationary equilibrium price jump or price mean return; $V_T$, $A_T$, and $E_T$ are total trading volume, reversal force, and liquidity energy on $T$ trading day, respectively.

Therefore, we have the price mean return $\bar{r}$ by a stationary equilibrium price jump and change in total trading volume $\Delta V$ in two consecutive trading days as

$$\bar{r}=\frac{\Delta p}{p_0}, \qquad (43)$$

and

$$\Delta V=\frac{V_T-V}{V} \qquad (44)$$

respectively. Obviously, $\bar{r}$ and $\Delta V$ could be positive, negative, or zero. Equations (43) and (44) are equations (3) and (4).

Equation (41) is equivalent to a joint two equations (40) and (42). The joint equations (40) and (42) are market crowd's adaptive trading price model. Its initial conditions are a stationary equilibrium price, total trading volume over a trading price range, and eigenvalue constant (agreement force in interaction) on *T-1* day. We can solve a joint two equations with three variables in mathematics.

In fact, there is no need to solve the joint two equations (40) and (42). We gather total trading volume from public database and determine a stationary equilibrium price on each trading day from test results. We figure out a stationary equilibrium price jump or price mean return by equation (43) and total trading volume increase or decrease by equation (44). Equations (40) and (42) provide dynamic mechanism how market environment results in a large supply and demand imbalance and a price mean return, and how market crowd adapt to gain and loss by trading volume.

## 6. DISCUSSIONS

In this section, we are going to explain trading volume implications and excessive trading, reply some questions from readers and discussants, and find potential applications in this research.

### 6.1 Trading Volume Implications and Excessive Trading

Trading volume has a lot of information about traders and environment. We highlight a number of interactions between Shi's price-volume differential equation and behavioral finance. The volume probability in the equation represents the frequency of market crowd's trading action. The behavioral annotation is a key link between market dynamics and crowd's trading behaviors in stock market.

In quantitative aspect, a trading volume distribution over a price range shows stationary equilibrium in crowd's interactive trading action or game. A stationary equilibrium price is the price on a trading day at which the corresponding volume is the maximal. In addition, trading volume in a time interval is trading action momentum or price momentum (see equation (24) and read Section 5.3.2). It could represent momentum force in equation (25), reversal force in equations (16), interactive or competitive force in equation (17), and agreement force in equation (38). The extremal eigenfunctions (35) and (39), which are obtained by solving Shi's equation, are optimal outcomes of crowd's trading action in which a trader could be rational, boundedly rational, or irrational, but they behave boundedly rational in general[41]. The eigenfunctions (35) and (39) show crowd's preference over a price range in trading action and the degree of agreement in crowd's interaction and competition if interactive force is a constant in equation (38). They have a set of an infinite number of discrete extremal values at a time.

In descriptive aspect, people trade a stock both in heuristics in terms of information and news (template learning) and in adaptation to outcome (reinforced learning). Judgment in both ways in uncertainty is convenient and useful for traders, but causes their cognitive biases and leads to severe and systematic errors. The trading action causes participants to trade more in a feedback loop with attitude to risks and sentiment such as overreaction, overconfidence, fear, greed, and regret etc. Adaptive learning by return outcomes increases trading volume. There is tremendous resistance to stop trading.

---

[41] Although seemly independent individuals often violate consistency and coherence because of decision frame (Tversky and Kahneman, 1981), market crowd behave coherence or agreement on a price in interaction and competition among themselves on a trading day (Shi, 2006).

As we have already reviewed the literature in introduction, there is a long list of biological and environmental factors that produces excessive trading volume. Therefore, it is probably a better and more complete explanation that interaction between information and news, the trading action, and return outcomes in a feedback loop produces excessive trading volume (see Figure 2). It includes various internal and external causes. The trading volume includes information about both interaction among trader's themselves and interaction between traders and environment.

## 6.2 Questions and Answers

What is major difference between stationary equilibrium in your study and equilibrium in traditional economics and finance?

A stationary equilibrium price is a crowd's agreement price in a time interval. It is an outcome of prior interactive and competitive trading action, whereas an equilibrium price is calculated in terms of information and news or fundamental value in neoclassical economics and finance. In addition, a stationary equilibrium price jumps from time to time (Kahneman and Tversky, 1979; Shi, 2006) on a trading day. It is dynamic, contrary to a static equilibrium price or "law of one price" in traditional economics and finance.

Had met crowd's own satisfying feelings, a trading outcome, an agreement price, or a stock value has already included private information value (Kyle, 1985), speculative value (Hirshleifer, Subrahmanyam, and Titman, 2006; Han and Kumar, 2013), sentiment value (Barberis, Shleifer, and Vishny, 1998; Baker and Wurgler, 2006; Han, 2007), attention value (Da, Engelberg, and Gao, 2011), gamble value (Kumar, 2009), entertainment value (Dorn and Sengmueller, 2009; Hoffmann and Shefrin, 2012), and so on. Therefore, a stock may have a set of an infinite number of discrete values or prices at a time. We could model them by the extremal eigenfunctions, equations (35) and (39). The model(s) has a set of an infinite number of discrete extremal values. For example, we could explain the segmented dual-class shares of firms in the study of Jia, Wang, and Xiong (2013).

A skeptical reader may argue that the trading action may disappear over time in a certain circumstance.

A three-term relationship in the trading action is: information and news, the trading action, and return outcome. There is the trading action as long as there is trading.

In general, market crowd behave both disposition effect in stock selling and herd behavior in stock buying simultaneously. They adapt to gain by trading volume increase and to loss by trading volume decrease. If disposition and herd effects disappear over time in a certain occasion, other crowd's trading behaviors might emerge significantly, for examples, greed holding and panic selling. Therefore, it disappears in a certain occasion a kind of crowd's trading behaviors rather than the trading action in stock market.

### 6.3 Potential Applications and Practices

There are many possible applications and practices in our study. First, a simple behavior is one of elements of a complex one. Behavior analysis and psychology will become a basic course for students in economics and finance in a long-term, similar to that mathematics and physics are basic courses for students in electronic engineering. Second, I think that the relationship between variables such as volume distribution over a price range has various applications. It could be used to build a quantitative supply and demand law in economics, a new stationary equilibrium theory in finance, and crowd's interactive game or competitive election theory in sociology. It is a new mathematical method for quantitative behavioral finance. Third, it helps us to understand market behaviors, for examples, excessive price volatility (Shiller, 1981; Leroy and Porter, 1981), the puzzles of equity premium (Mehta and Prescott, 1985), and asset bubble (Freshen, Gateman, and Rouwenborst, 2009; Soros, 2010) etc. Fourth, smart and sophisticated investors are able to buy when a stock value is much under-estimated, and sell when it is highly over-estimated in a long-term, because asset prices or outcomes include both fundamental value and behavioral values. Fifth, speculative traders and market makers do have a lot of opportunity to earn abnormal return because there are multi-possible discrete equilibrium points or asset prices of a stock. For example, an equilibrium point or asset price of a small size stock is relatively easy to be changed in uncertain environment. Sixth, we recommend taking different trading strategies and investment portfolios with change in environment because market and crowd's behaviors vary with time and environment. For example, return is higher but risk is lower when stock market is bullish. Seventh, it would be better to invest index if one could not predict the outcome of individual stocks because people adapt to outcome efficiently on a daily basis. Eighth, it might be possible for us to develop a feasible automated trading strategy and run automated, algorithmic, and high frequency trading in financial market in terms of Shi's price-volume probability wave equation and market crowd's trading behaviors if outcome arbitrage could cover the cost involved, because stationary equilibrium exists widely on a trading day in financial market, particularly in bond market (Shi, 2013).

### 7. CONCLUSIONS, CONTRIBUTIONS, AND EXTENSIONS

Adaptive learning is seen everywhere in economics and finance. Guided by behavior analysis and a price-volume differential equation (Shi, 2006), we measure the frequency of the trading action by trading volume probability, study crowd's trading behaviors, and test two kinds of adaptive trading behaviors relevant to the volume uncertainty associated with price in stock market. With prospect theory, we conclude that: 1) market crowd trade a stock not only in simple heuristics but also in efficient adaptation; they gradually tend to achieve agreement on an outcome or an asset price widely on a trading day and generate such a stationary equilibrium price very often in interaction and competition among themselves no matter whether it is highly overestimated or underestimated; they trade a stock effectively in reference to an outcome instead of the fundamental value that is hardly calculated; 2) market crowd adapt to gain and loss by trading volume increase or decrease significantly in interaction with environment in any two consecutive trading days; 3) Considering more gains (more losses) against gains (losses), market crowd adapt to trade less (can't help to trade more in fear of more losses) when they are consistently rewarded in bull market (they are severely punished in surprise by abrupt and deep price drop). While the effect of herd buyers and disposition sellers disappears, other trading behavioral anomalies such greed and fear might pronounce significantly. Market crowd trading behavioral "anomalies" vary in response to gains and losses by trading volume increase or decrease over time and occasion.

We provide a new conceptual framework and make several contributions to economics and finance. First, trading volume probability has a lot of information about traders and market. It is a key link between market dynamics and crowd's trading behaviors. Second, the interaction between information and news, the trading action, and return outcome in the three-term feedback loop produces excessive trading volume. It includes a variety of internal and external causes for excessive trading. Third, adaptation to outcome (reinforced learning), we test, is the other judgment under uncertainty which leads cognitive biases, except for heuristics in terms of information and news (template learning). Adaptation and heuristics are insensitive each other (Kahneman and Tversky, 1973). We use them to explain underlying trading behaviors in a liquidity utility hypothesis, equation (13). Fourth, a stationary equilibrium price is an outcome of crowd's trading action. It jumps from time to time, contrary to a static equilibrium price in traditional economics and finance. Had met crowd's own satisfying feelings, fifth, an asset price, the outcome of crowd's trading action, has already included private information, speculative, sentiment, attention, gamble, and entertainment values etc. Therefore, asset prices for a firm at a time could be a set of an infinite number of discrete prices in real world instead of one price in neoclassical finance. Explicit models, equations (35) and (39), capture this characteristic with an infinite number of discrete extremal values. We are able to extend this paper, furthermore, to write that maximizing an expected utility does not correspond to investor's rationality. An expected utility maximizer which generates an infinite number of extramal values is applied to bounded rationality rather than rationality. Moreover, a liquidity utility hypothesis, the least price variation principle, and the price-volume probability differential equation are new methods in mathematical finance. We describe the dynamic behavior of a stock, which has multi-possible discrete equilibrium points or asset prices, by new mathematical finance. With insights on crowd's trading behaviors and new mathematical models, we might fully understand how asset prices are determined and how mispricings of assets emerge. There are many potential applications and practices in sociology, economics, and finance in the future.

**Supplement A: Eigenvalue and Probability Wave**

Let us have brief explanation on eigenvalue and probability wave for readers who are not majored in physics.

An eigenvalue is an observable and measurable constant in a distribution function whereas a parameter is an unknown constant. Thus, it is possible for us to test an eigenfunction (a distribution function with an eigenvalue) and understand its mechanism by its observable(s). For example, there are two sets of eigenvalue in trading volume distribution over a price range. One is expressed by equation (28), by which we analyze how liquidity utility constrains market crowd behavior and explain why they could behave independently in some particular cases. The other is expressed by equations (31) or (32), by which we find that market crowd behave coherently between momentum traders and reversal traders in interaction among themselves. They reach agreement on a stationary equilibrium price at which the corresponding volume achieves a maximal. However, we do not have what information is in a parameter in a distribution function.

Probability wave is a kind of wave in which we use volume probability rather than the amplitude of a wave to describe its intensity. For example, we measure the intensity of price volatility by trading volume probability rather than price volatility return in a time interval (reference to Figure S-1).

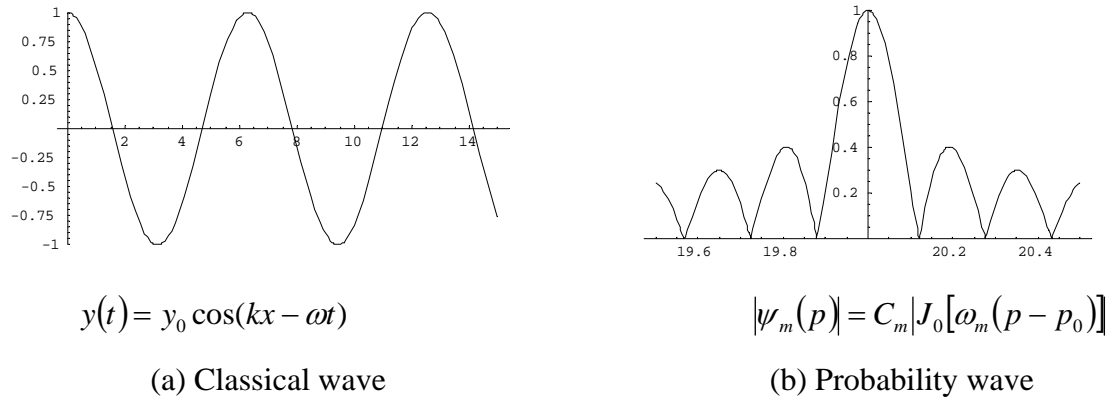

$y(t) = y_0 \cos(kx - \omega t)$ $\qquad\qquad |\psi_m(p)| = C_m |J_0[\omega_m(p - p_0)]|$

(a) Classical wave  (b) Probability wave

Figure S-1: Classical wave and probability wave

Now, let us compare and contrast a probability wave with a classical wave. First, x-coordinate stands for price and y-coordinate denotes trading volume probability in a time interval in the price-volume probability wave, while x-coordinate and y-coordinate represent time and the amplitude of a wave in a classical wave, respectively. Second, we describe the intensity of a probability wave by its volume probability, whereas we describe the intensity of a classical wave by its amplitude. Third, the intensity of a probability wave is equal to and larger than zero, while that of a classical wave could be positive, negative, and zero. Forth, the intensity of a probability wave shows a wave change with an independent variable. Larger amplitude is not equal to stronger intensity in a probability wave. For example, the intensity of a larger price volatility could be very weak if there is a very little of trading volume at a price. In a classical wave, the larger the amplitude is, the stronger the intensity does be. Fifth, there could be no phase position in a probability wave, contrary to that there is phase position in a classical and plane wave. A probability wave actually describes a kind of distribution such as many-body or crowd behavior.

However, there are some commons in both. For examples, they behave coherently and show a repeated change in reference to an equilibrium point. Both can characterize positive and negative correlation between two variables (see figure S-1).

**Supplement B: Data, Data Process, and Empirical Tests**

1. Data and Data Mining

We use tick by tick high frequency data in Huaxia SSE 50ETF (510050) on a daily basis in China stock market from April 2, 2007 to April 10, 2009, more than two years when it experiences a whole course from bubble growth, burst, shrink to reversal again. There are about 740 days and 495 trading days in the two year term, i.e., there are 495 distributions in our tests. The data is from HF2 database in Harvest Fund Management Co., Ltd.

We process the data in two steps. First, we reserve two places of decimals in price by rounding-off method and add the volume at a corresponding price[42]. Second, accumulative trading volume at a price is divided by total volume in a trading day. Thus, we obtain the volume probability at the price and a volume distribution over a trading price range in a trading day.

2. Regression and Significant Tests

When market crowd reach agreement on a stationary equilibrium price, we have a theoretical agreement trading price model as

$$|\psi_m(p)| = C_m |J_0[\omega_m(p - p_0)]|, \qquad (m = 0,1,2\cdots) \qquad \text{(S-1)}$$

where $C_m$, $\omega_m$, and $p_0$ are a normalized constant, an eigenvalue constant, and a stationary equilibrium, respectively. They are three constant coefficients and can be determined by its nonlinear regression model,

$$|\psi_{m,i}(p_i)| = C_m |J_{0,i}[\omega_m(p_i - p_0)]| + \varepsilon_i, \qquad (i = 1,2,3\cdots,n) \qquad \text{(S-2)}$$

where $n$ is the number of prices over a trading price range in a trading day; $\varepsilon_i$ is random error subject to $N(0,\sigma^2)$; $|\psi_{m,i}(p_i)|$ is an observed accumulative trading volume probability at a price, while $C_m |J_{0,i}[\omega_m(p_i - p_0)]|$ is a theoretical accumulative trading volume probability.

We test volume distribution over a trading price range by regression model, equation (S-2), using Origin 6.0 Professional software in which Levenberg-Marquardt nonlinear least square method is used. By doing so, we fit the distributions and get the values of $C_m$, $\omega_m$ and $p_0$ from test reports (see (a) in Figure S-2).

We test significance by $F$ statistic. The coefficient of determination $R^2$,

$$R^2 = \frac{ESS}{TSS} = \frac{TSS - RSS}{TSS}, \qquad \text{(S-3)}$$

---

[42] Original data reserves three places of decimals.

where $ESS = \sum_{i=1}^{n}(\hat{Y}_i - \overline{Y})^2$, $RSS = \sum_{i=1}^{n}(Y_i - \hat{Y}_i)^2$, and $TSS = \sum_{i=1}^{n}(Y_i - \overline{Y})^2$ are the explained sum of squares, the residual sum of squares, and total sum of squares, respectively.

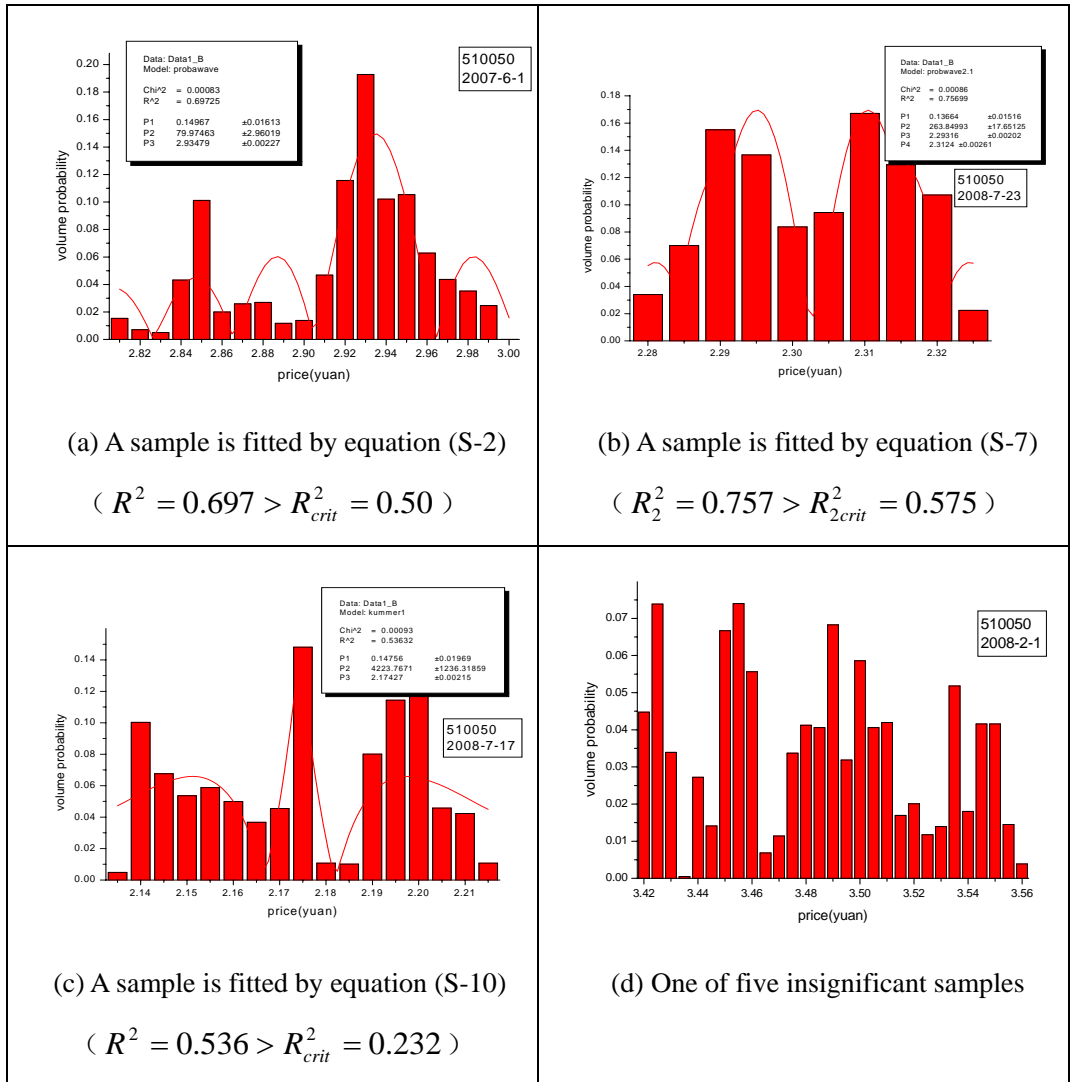

(a) A sample is fitted by equation (S-2)

($R^2 = 0.697 > R^2_{crit} = 0.50$)

(b) A sample is fitted by equation (S-7)

($R^2_2 = 0.757 > R^2_{2crit} = 0.575$)

(c) A sample is fitted by equation (S-10)

($R^2 = 0.536 > R^2_{crit} = 0.232$)

(d) One of five insignificant samples

Figure S-2: The volume distribution test reports in samples [43]

We have

$$F = \frac{ESS/k}{RSS/(n-k-1)}, \qquad (S-4)$$

where $n$ and $k$ are sample size and the number of explanatory variables, respectively. If $F > F_{0.05}$ or

$$R^2 > R_{crit}^2 = \frac{k \cdot F_{0.05}}{k \cdot F_{0.05} + (n-k-1)}, \qquad (S\text{-}5)$$

then, the test holds true at 95% significant level. Here, $k=1$.

---

[43] In Figure 5, P1, P2, and P3 are a normalized constant, an eigenvalue, and a stationary equilibrium price, respectively. P4 is a stationary equilibrium price, too.

Our test result is that 380 out of 495 distributions (about 76.77%) show significance (see (a) in Figure S-2). The remainders (about 23.23%) lack significance.

There are two notable characteristics among the distributions without significance. First, the number of trading prices is fewer and the sample size is not large enough over prices for statistic test. It is partly credited by previous data process, in which we reserved two places of decimals in price by rounding-off three places in original data. The data process results in information loss.

To solve the problem, we add 0.005 in three places of decimals in price and subdivided volume at corresponding prices. Then, we fit and test the remainders by equations (S-2) and (S-5) again. 28 more distributions show significance. Thus, there are total 408 (about 82.42%) distributions showing significance (see Table Ⅰ). We get stationary equilibrium and agreement prices in these trading days from our test reports.

Second, 87 remainders have at least two volume kurtosis over a trading price range. It is explained that there is stationary equilibrium price jump on these trading days. Trading price is volatile upward and downward from one stationary equilibrium price to another. We can model them by the linear superposition of equation (S-2), that is,

$$|\psi_m(p)| = \sum_n C_m |J_0[\omega_{m,n}(p - p_{0n})]|, \qquad (n=1,2\cdots) \quad (S\text{-}6)$$

where $n$ is the number of stationary equilibrium prices. We choose $n=2$, fit and test two of volume kurtosis in distribution by following regression model,

$$|\psi_{m,i}(p_i)| = \sum_i \sum_{n=1,2} C_m |J_{0,i}[\omega_{m,n}(p_i - p_{0,n})]| + \varepsilon_i \qquad (i=1,2\cdots) \quad (S\text{-}7)$$

where $n=2$.

Table Ⅰ: Test Reports on Stationary Equilibrium and Agreement Price(s)

|  | No. of Distributions | Percentage (%) |
| --- | --- | --- |
| Total Number of Distributions | 495 | 100 |
| Agreement and Stationary Equilibrium | 408 | 82.42 |
| Agreement on Two Prices (Jump and Adaptation) | 59 | 11.92 |
| Agreement on Three Prices (No Agreement) | 23 | 4.65 |
| Uniform Agreement (No Agreement) | 5 | 1.01 |

Note: Trading time interval is 4 hours per day in China stock market.

In the tests, $k=2$ and $R_2^2 > R_{2crit}^2$. Our test result is: 59 (11.92% in total) show significance at 95% level among 87 distributions. Market crowd adapt and have agreement on two prices (see (b) in Figure S-2 and Table Ⅰ).

For the rest of 28 distributions, we use market crowd independent trading price model,

$$\psi_{n,m}(p) = C_{n,m} e^{-\sqrt{A_{n,m}}|p-p_0|} \cdot F\left(-n, 1, 2\sqrt{A_{n,m}}|p-p_0|\right). (n, m = 0,1,2,\cdots) \quad \text{(S-8)}$$

In convenience, we choose $n = 1$. It is

$$\psi_{1,m}(p) = C_{1,m} e^{-\sqrt{A_{1,m}}|p-p_0|} \cdot F\left(-1, 1, 2\sqrt{A_{1,m}}|p-p_0|\right)$$

$$= C_{1,m} e^{-\sqrt{A_{1,m}}|p-p_0|} \cdot \left\|1 - 2\sqrt{A_{1,m}}|p-p_0|\right\| , \quad \text{(S-9)}$$

and its regression model is

$$\left|\psi_{1,m,i}(p_i)\right| = C_{1,m} e^{-\sqrt{A_{1,m}}|p_i-p_0|} \cdot \left|F_i\left(-1,1,2\sqrt{A_{1,m}}|p_i-p_0|\right)\right| + \varepsilon_i, (i = 1,2\cdots) \text{(S-10)}$$

23 distributions (about 4.65% in total) show significance at 95% level, modeled by equation (S-10). Please see (c) in Figure S-2. The rest 5 distributions still lack significance (see (d) in Figure S-2). Market crowd tend to trade uniformly and behave no agreement on a price in the 28 distributions (5.66% in total).

The empirical test results are listed in Table Ⅰ, consistent with those by Shi (2006).

### 3. Correlation and Significant Test

In behavior analysis, we usually study observable correlations between an objective event which increases the frequency of the response and an observable behavior upon which it is contingent[44]. Then, we explain the underlying thought and expectation on reinforcement. Now, we make correlation analysis on a daily basis over a two-year period between price volatility return and the frequency of market crowd trading action, that is, the intensity of market crowd trading conditioning. It provides a close look how past price volatility return influences market crowd current trading action over time and occasion.

There are 408 accumulative volume distributions (about 82.42% in total) that show significance in our test by the absolute of zero-order Bessel eigenfunction regression model. We can get the stationary equilibrium prices from test reports directly. For the rest (87 distributions), we choose the volume weight price mean values. In this way, we can figure out the rate of mean return in any two consecutive trading days and study the correlation between the rate of mean return and the change in the intensity of trading conditioning. Here, the change in the intensity of trading conditioning is approximately equal to the rate of change in total trading volume in two days. It is determined by equation (31).

Correlation coefficient $r_{X,Y}$ is given by

$$r_{X,Y} = \frac{\text{cov}(X,Y)}{\sigma_X \sigma_Y}, \quad \text{(S-11)}$$

---
[44] Feedback reinforcement may be seen as a discriminative stimulus.

where $\sigma_X$ and $\sigma_Y$ are the standard deviations of variable $X$ and $Y$, $\text{cov}(X,Y)$ is covariance. We use $t$-statistic to test significance. If we have

$$H_0: \rho = 0, \quad H_1: \rho \neq 0, \tag{S-12}$$

then,

$$t = \frac{|r - \rho|}{\sqrt{(1-r^2)/(n-2)}}, \tag{S-13}$$

where $r$ and $n$ are correlation coefficient and sample size, respectively. For $\alpha = 0.05$, if $t > t_{crit} = t_{0.05/2}(n-2)$, then, original hypothesis is rejected. Correlation coefficient is significant not equal to zero at 95% level.

Table II : Test Reports on Correlation and Its Significance

| | Terms | Number of Distributions | SSE Composite Index (1A0001) | Correlation Coefficients and Its Significant Test Results |
|---|---|---|---|---|
| A | 2007.4.2—2009.4.10 | 494 | 3252.59—2444.23 | 0.1391 (t=3.115>t$_{crit}$=1.960) |
| B | 2007.4.2—2007.6.29 | 59 | 3252.59—3820.70 | -0.2567 (t=2.006> t$_{crit}$ =2.001) |
| C | 2007.7.2—2007.10.30 | 83 | 3836.29—5954.77 | **0.0729** **(t=0.6583< t$_{crit}$ =1.990)** |
| D | 2007.11.1—2008.4.30 | 122 | 5914.28—3693.11 | **0.1026** **(t=1.130< t$_{crit}$ =1.980)** |
| E | 2008.5.5—2008.10.31 | 123 | 3761.01—1728.79 | 0.1963 (t=2.202> t$_{crit}$ =1.980) |
| F | 2008.11.3—2009.4.10 | 107 | 1719.77—2444.23 | 0.4766 (t=5.556> t$_{crit}$ =1.983) |

Notes:
1) Correlation specifies the correlation between the rate of mean return and the change in the intensity of trading conditioning in any two consecutive trading days;
2) Here, $t_{crit}$ is $t_{0.05/2}(n-2)$; If $t > t_{crit}$, then, the correlation coefficient is significantly not equal to zero; On the other hand, we can not reject original hypothesis that the correlation coefficient is equal to zero;
3) It is printed in bold if test result does not show significance;
4) SSE Composite Index is measured by closing point.

We subdivide two year high frequency data into 5 time intervals from bubble growth, burst, and shrink until market reversal again in China, a whole course that is paralleled with the growth, burst, and collapse of sub-prime bubble that originated in the United States in 2008 and set off a chain reaction worldwide (reference to Table Ⅱ). The first term is from April 2, 2007 (SSE Composite Index at 3252.59 points) to June 29, 2007 (SSE Composite Index at 3820.70 points), the first half before bubble burst in China. The second is from July 2, 2007 (SSE Composite Index at 3836.29 points) to October 31, 2007 (SSE Composite Index at 5954.77 points), the second half before bubble burst in China. The third is from November 1, 2007 (SSE Composite Index at 5954.77 points) to April 40, 2008 (SSE Composite Index at 3693.11 points), the first half after bubble burst in China. The forth is from May 5, 2008 (SSE Composite Index at 3761.01 points) to October 31, 2008 (SSE Composite Index at 1728.79 points), the second half after bubble burst in China. And the last is from November 3, 2008 (SSE Composite Index at 1719.77 points) to April 10, 2009 (SSE Composite Index at 2444.23 points), price reversal time interval after a year deep drop (reference to TableⅡ).

There is advantage in this approach: we can study market crowd behaviors in learning over time and environment.

In our test, we use Eviews 6.0. We have several main findings. First, there is significant positive correlation in general between the rate of mean return and the change in the intensity of trading conditioning (reference to line A in TableⅡ). Second, correlation coefficient varies in 5 subdivided periods. For examples, (a) they lack significance in spite of positive correlations in two time intervals right before and just after bubble burst (reference to line C and D in TableⅡ); (b) it shows positive significance in the second half after bubble burst; (c) there is the highest positive correlation during price reversal time interval after a year deep drop. Its correlation coefficient is 0.4766 (reference to line F in TableⅡ); (d) particularly, there exists significant negative correlation (correlation coefficient is -0.2567) when SEE Composite Index is rising during bull market (reference to line B in TableⅡ). We discuss them in section 5.